\documentclass[prb, twocolumn, superscriptaddress, showpacs]{revtex4}
\usepackage{tabularx,graphicx,subfigure,float}
\usepackage{dcolumn}
\usepackage{bm}
\usepackage{color}

\begin{document}

\title{First-Principles Equation of State Calculations of Warm Dense Nitrogen}

\author{K. P. Driver}
 \affiliation{Department of Earth and Planetary Science, University of California, Berkeley, California 94720, USA}
 \email{kdriver@berkeley.edu}
 \homepage{http://militzer.berkeley.edu/~driver/}

\author{B. Militzer}
 \affiliation{Department of Earth and Planetary Science, University of California, Berkeley, California 94720, USA}
 \affiliation{Department of Astronomy, University of California, Berkeley, California 94720, USA}

\date{\today}

\begin{abstract}
Using path integral Monte Carlo (PIMC) and density functional
molecular dynamics (DFT-MD) simulation methods, we compute a coherent
equation of state (EOS) of nitrogen that spans the liquid, warm dense
matter (WDM), and plasma regimes. Simulations cover a wide range of
density-temperature space, $1.5-13.9$~g$~$cm$^{-3}$ and $10^3-10^9$~K.
In the molecular dissociation regime, we extend the
pressure-temperature phase diagram beyond previous studies, providing
dissociation and Hugoniot curves in good agreement with experiments
and previous DFT-MD work.  Analysis of pair-correlation functions and
the electronic density of states in the WDM regime reveals an evolving
plasma structure and ionization process that is driven by temperature
and pressure.  Our Hugoniot curves display a sharp change in slope in the
dissociation regime and feature two compression maxima as the K and
L shells are ionized in the WDM regime, which have some significant
differences from the predictions of plasma models.
\end{abstract}



\maketitle

\section{INTRODUCTION}

Nitrogen is a prototypical molecular system known for its large
cosmological abundance, ability to form numerous chemical compounds,
and its interesting solid, liquid, and electronic phase transitions at
high pressures and temperatures~\cite{Ross2000,Nellis2002}.  Nitrogen
can be found over a wide range of physical and chemical conditions
throughout the universe, ranging from low densities in interstellar
space~\cite{Meyer1997} to extreme densities in stellar
cores~\cite{Wallerstein1997}, and it plays important roles in
planetary atmospheres~\cite{Lodders2002} and interiors of ice giant
planets~\cite{hubbard_planets}.  In the condensed matter regime,
nitrogen is capable of forming a wide variety of triple-, double-, and
single-bonded compounds, which makes it of interest to geological and
energy sciences.  At higher densities and temperatures, nitrogen
exists as a molecular fluid that undergoes a pressure-induced
dissociation transition to polymeric and atomic fluids of interest in
planetary science~\cite{Nellis1991,Ross2006,Boates2009}.  In the warm
dense matter (WDM) regime, nitrogen exists in partially ionized plasma
states, which are of fundamental interest to shock physics and
astrophysics communities.  An accurate understanding of the equation
of state (EOS) in these regimes is important for determining the
thermodynamic properties of the various nitrogen phases and their
implications for science and technology.

\begin{figure}
  \begin{center}
        \includegraphics*[width=8.6cm]{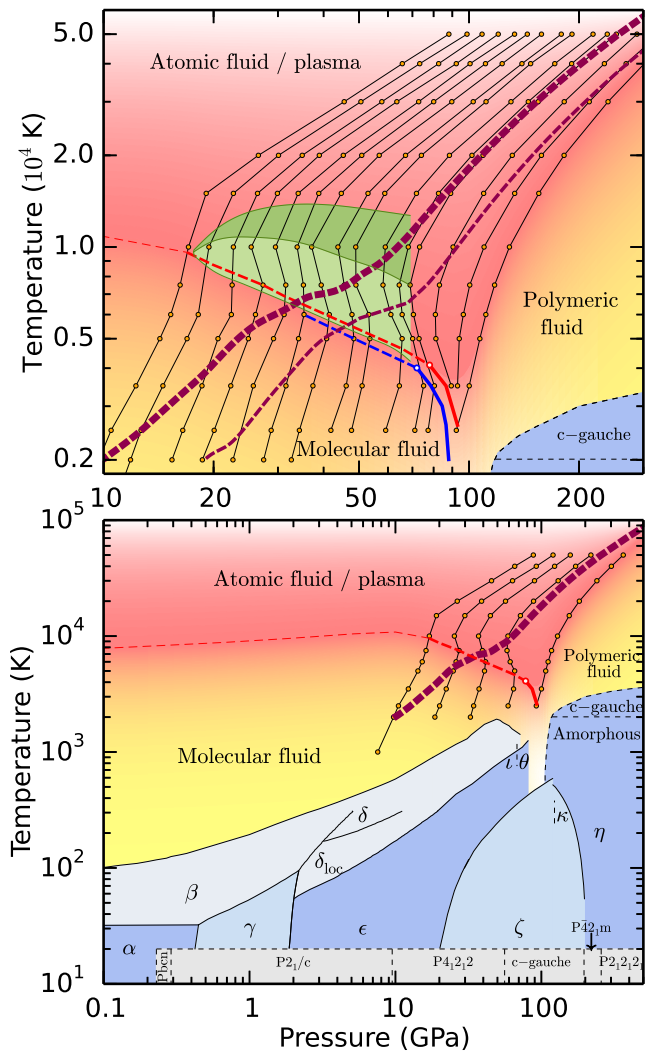}
  \end{center}

\caption{Pressure-temperature phase diagram of nitrogen.  The lower
  panel displays solid phases; molecular, polymeric, and atomic fluid
  phases; and the plasma regime. Phases well characterized by
  experiments are outlined with solid black lines, while others are
  outlined with a dashed line.  Circles represent a subset of our
  DFT-MD isochore data used to compute the Hugoniot (thick, short-dashed
  curve) and dissociation curves. The latter changes from a dashed to
  solid curve to indicate the change to a first order liquid-liquid
  transition region. The upper panel is a magnified view of the
  molecular dissociation region, showing a larger subset of our DFT-MD
  calculations. The thick and thin dashed curves are our predicted
  Hugoniot curves for two different initial densities of 0.808 and
  1.035 g~cm$^{-3}$, respectively. Here, we also compare our
  dissociation curve with previous DFT-MD simulations by
  Boates~\emph{et al.}~\cite{Boates2009} (blue line). The green shaded
  area marks the region from the onset of dissociation, where the
  isochores begin to show that $(\partial P/ \partial T)_V<0$, to the
  point at which pressure returns to its value before the onset of
  dissociation.}

  \label{fig:PTdiagram}
\end{figure}

At ambient conditions, nitrogen exists as a diatomic gas comprised of
strong, triply-bonded dimers. At low T, nitrogen forms a molecular
solid that undergoes a series of solid phase transitions with
increasing pressure (see Fig.~\ref{fig:PTdiagram}), which have been
identified by a number of static compression
experiments~\cite{Reichlin1985,Mills1986,Olijnyk1990,Bini2000,Gregoryanz2002,Gregoryanz2007}.
Around 50-70 GPa, density functional molecular dynamics (DFT-MD)
simulations first
predicted~\cite{McMahan1985,Martin1986,Mailhiot1992,Katzke2008,Pickard2009,Erba2011}
the triple bond would destabilize to form various lower-energy,
nonmolecular (possibly amorphous), polymeric phases composed of
double- or single-bonded atoms, such as cubic
gauche~\cite{Mailhiot1992}. Later, static compression experiments
confirmed the transition to nonmolecular
phases~\cite{Goncharov2000,Eremets2001,Gregoryanz2001,Gregoryanz2002,Eremets2004natmat,Eremets2004JCP,Popov2005,Lipp2007,Eremets2007,Gregoryanz2007,Chen2008}.
The most extreme static compression experiments thus far have measured
the equation of state up to a pressure of 270
GPa~\cite{Gregoryanz2001} and temperatures ranging up to 2,000
K~\cite{Gregoryanz2007}.  First-principles
simulations~\cite{Mailhiot1992,Mattson2004,Zahariev2005,Zahariev2006,Wang2007,Yao2008,Ma2009,Pickard2009,Wang2010,Boates2011,Erba2011,Wang2012}
have predicted solid molecular and nonmolecular phases up to pressures
as high as 400 GPa~\cite{Pickard2009}. However, first-principles
predictions do not agree with experiments on what high pressure phases
are stable at $T=0$~K (Fig.~\ref{fig:PTdiagram}), which continues to
make solid nitrogen an interesting test case for improved experimental
and theoretical methods.

While the solid phases have been intensely studied, the liquid, and,
particularly the WDM and plasma states, have been investigated to a
lesser extent.  In this work, we focus on extending the studies of the
EOS of liquid, WDM, and plasma states of nitrogen (Fig. 2).  Several
experimental measurements of dense, liquid nitrogen states have been
performed using dynamic shock compression
experiments~\cite{Zubarev1962,Dick1970,Nellis1980,Nellis1984,Schott1985,Radousky1986,Moore1989,Nellis1991,Chau2003},
with the most extreme ones reaching up to a pressure of 180
GPa~\cite{Chau2003} and a temperature of 14,000 K~\cite{Nellis1991}.
The main focus of these experiments was to understand the
shock-induced dissociation of molecular nitrogen at 30-80 GPa on the
Hugoniot curve, as reviewed by Ross~\cite{Ross2000} and
Nellis~\cite{Nellis2002}.  Nitrogen is also particularly interesting
among the diatomic molecules because it exhibits unexpected phenomena,
such as reflected-shock-induced cooling, where the dissociation to
a polymeric fluid gives rise to a region of the phase diagram with
$(\partial P/ \partial T)_V<0$ (Fig.~\ref{fig:PTdiagram}). In this
work, we revisit the dissociation curve and connect the liquid EOS to
the WDM and plasma regimes.

Theoretical studies of shock-induced dissociation of dense, fluid
nitrogen have been performed with a variety of approaches. A number of
semi-empirical techniques have been employed, such as fluid
variational
theory~\cite{Ross1980,Schott1985,Ross1987,Ross1992,Ross2006,Chen2006},
molecular dynamics~\cite{Johnson1983}, chemical equilibrium
models~\cite{Kerley1986N,Hamilton1989,Thiel1996}, Monte
Carlo~\cite{Belak1988}, and integral equation
theory~\cite{Fried1998}. First-principles DFT-MD has also been used to
study shocked, fluid states~\cite{Kress2000,Mazevet2001,Mattson2011}.
The semi-empirical and DFT-MD studies have been successful in
predicting the principal Hugoniot curve and doubly shocked cooling
within the dense, fluid dissociation regime (up to 110 GPa and 20,600
K)~\cite{Kress2000}, in good agreement with the experimental
measurements. In addition, Ross and Rogers~\cite{Ross2006} have used
the activity expansion method (ACTEX)~\cite{Rogers1997} to compute the
Hugoniot curve in the plasma regime. ACTEX is a semi-analytic plasma
model parameterized by spectroscopic data and is based on the grand
partition function for a Coulomb gas of ions and electrons. It has
been successful at predicting plasma properties in the
weak-to-moderate coupling regime~\cite{Rozsnyai2001}. The ACTEX model
identifies a Hugoniot curve compression maximum associated with K shell (1s)
ionization, which will be discussed in more detail in Section VI.

\begin{figure}[t]
  \begin{center}
        \includegraphics*[width=8.6cm]{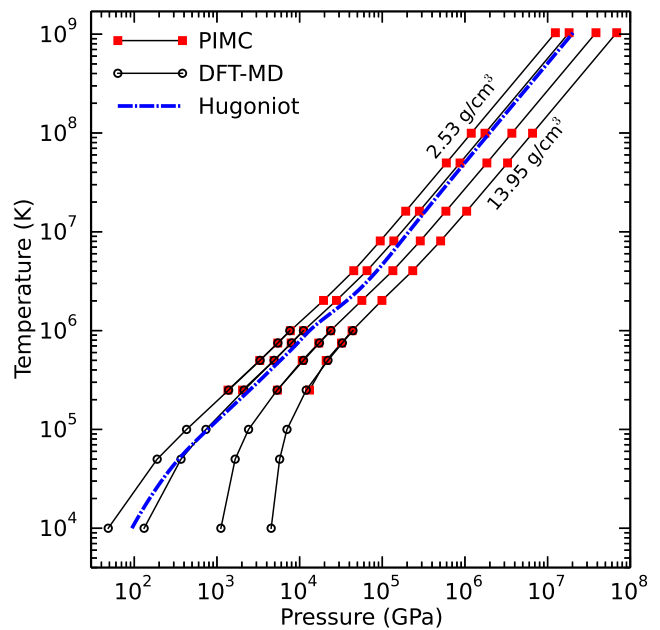}
  \end{center}

    \caption{Temperature-pressure isochores computed with DFT-MD
      (circles) and PIMC (squares) at densities of 2.5, 3.7, 7.8, and
      13.9 g~cm$^{-3}$. The blue dash-dotted line shows the Hugoniot
      curve for an initial density of $\rho_0 = 1.035$ g~cm$^{-3}$}

  \label{TvsP}
\end{figure}

DFT-MD has provided the most accurate description of liquid and warm
dense states of nitrogen up to moderate temperatures ($\sim$10$^5$
K). However, for higher temperature applications, such as
astrophysical modelling and exploring pathways to fusion, a
first-principles method that extends the EOS across the entire high
energy density physics regime, bridging the liquid, WDM, and plasma
regimes, is still needed. PIMC is one of the most promising
first-principle methods to extend our study beyond the scope of DFT-MD
because it is based on a quantum statistical many-body framework that
naturally incorporates temperature effects and, in addition, becomes
more efficient at higher temperatures.  Building on earlier
simulations of hydrogen~\cite{PC94,Ma96,Mi99,MC00,MC01,Mi01} and
helium~\cite{Mi06,Mi09,Mi09b}, we have been extending the PIMC
methodology for WDM composed of increasingly heavy
elements~\cite{Mi09, Driver2012, BenedictCarbon,
  Driver2015Oxygen,Driver2015Neon,Militzer2015Silicon}. Here, we apply
our PIMC and DFT-MD simulations to liquid and WDM states of nitrogen
over much wider density-temperature range (1.5--13.9 g$\,$cm$^{-3}$
and 10$^3$--10$^9$ K, see Figs. \ref{fig:PTdiagram} and \ref{TvsP})
than has been previously explored with DFT-MD alone.

The paper is organized as follows: In Section II, we describe PIMC and
DFT-MD simulation methods for liquid and WDM regimes. In Section III,
we first discuss the DFT-MD calculations of the liquid EOS, its
dissociation transition, and present an updated phase diagram. We then
extend the liquid EOS into the WDM and plasma regimes and show that
DFT-MD and PIMC produce consistent results for intermediate
temperatures.  In section IV, we characterize the structure of the
plasma and ionization processes by examining changes in different
pair-correlation functions as a function of temperature and
density. In section V, we discuss the electronic density of states to
provide further insight into the ionization process. In section VI, we
discuss shock Hugoniot curves. Finally, in section VII, we summarize
our findings.

\section{SIMULATION METHODS}

PIMC~\cite{Ce91,Ceperley1995,Ce96} is a state-of-the-art
first-principles method for computing the properties of interacting
quantum systems at finite temperature. Since PIMC is based on the
thermal density matrix formalism, it naturally incorporates
temperature into the framework. The density matrix is expressed in
terms of Feynman's imaginary time path integrals, which are evaluated by
efficient Monte Carlo techniques, treating electrons and nuclei
equally as quantum paths that evolve in imaginary time without
invoking the Born-Oppenheimer approximation. Therefore, PIMC is able
to explicitly treat all the effects of bonding, ionization,
exchange-correlation, and quantum degeneracy in a many-body framework
that simultaneously occur in the WDM regime~\cite{Koenig2005}. The
Coulomb interaction is incorporated via pair density matrices derived
from the eigenstates of the two-body Coulomb
problem~\cite{Pollock1988,MG06}.  The efficiency of PIMC increases
with temperature as particles behave more classically at higher
temperatures and fewer time slices are needed to describe quantum
mechanical many-body correlations.

PIMC requires a minimal number of controlled approximations, which are
minimized by converging the time step and system size. We determined
the necessary time step by converging total energy until it changed by
less than 1.0\%. We use a time step of 1/256 Ha$^{-1}$ for
temperatures below 4$\times10^6$ K.  For higher temperatures, we
decreased the time step as 1/T. In order to study finite size errors,
we perform simulations with 8 and 24 atoms in cubic simulations cells
and found that the total energy differed by 0.4\% or
less~\cite{Driver2015Neon}. All results for the internal energy and
pressure that we report have statistical errors of 0.3\% or less.

The only uncontrolled approximation in PIMC is the fixed-node
approximation that is introduced to avoid the fermion sign
problem~\cite{Ceperley1991}. We employ a free-particle nodal
structure, which we have shown to work reliably for partially ionized
hydrogen~\cite{MC01}, helium~\cite{Mi09}, carbon~\cite{Driver2012},
water~\cite{Driver2012}, oxygen~\cite{Driver2015Oxygen}, and
neon~\cite{Driver2015Neon}.  Free-particle nodes work well as long as
only a small number of bound electronic states are occupied. For
plasmas of first-row elements, we have found that free particle nodes
yield good results for conditions where the 1s states are fully
occupied and the 2s states are partially
occupied~\cite{Driver2012}. Lower temperature conditions can be
studied efficiently with DFT-MD.

\begin{figure}[t]
  \begin{center}
        \includegraphics*[width=8.6cm]{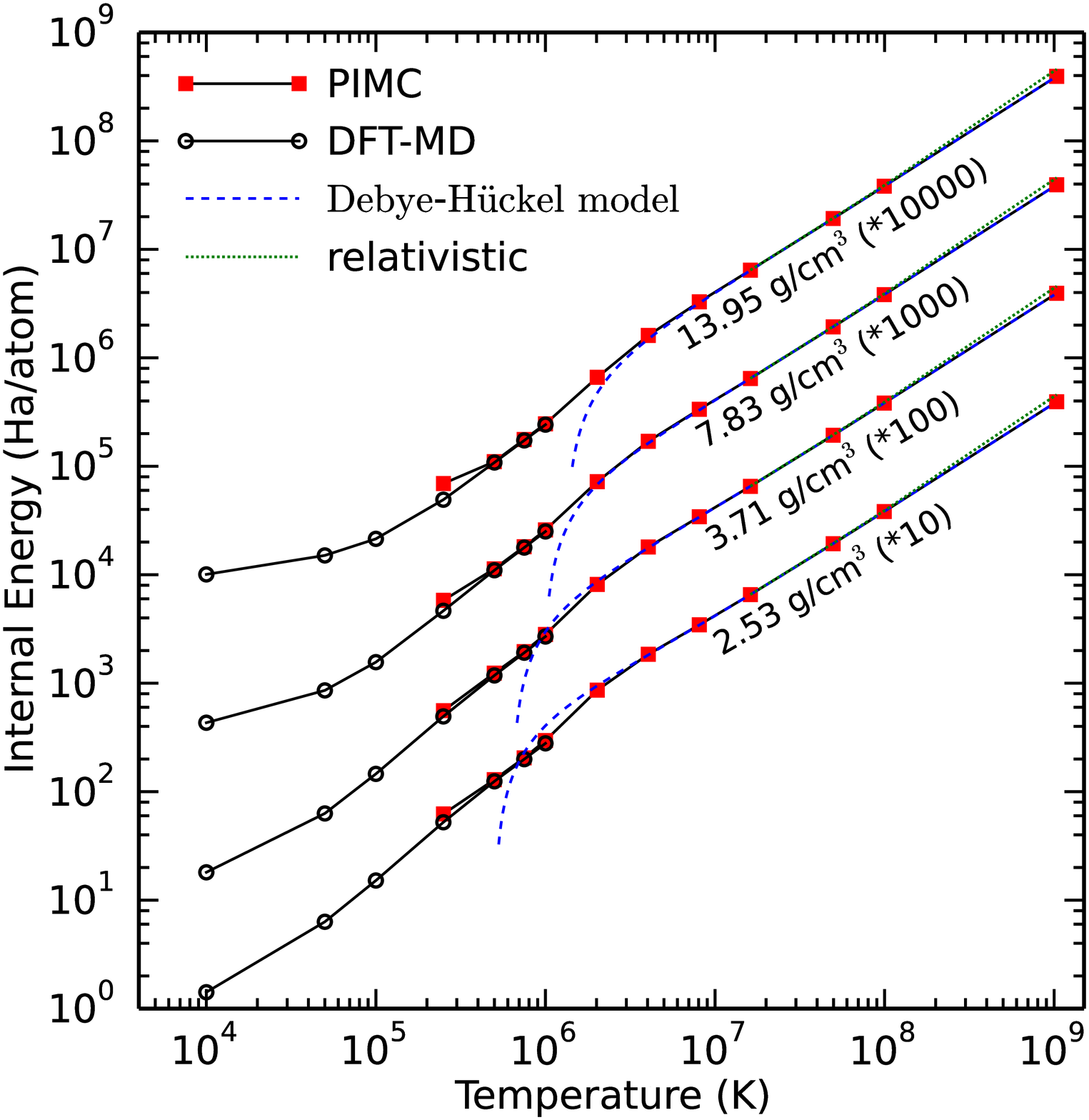}
  \end{center}

    \caption{Isochores computed with PIMC (squares), DFT-MD (circles),
      and the Debye-H\"{u}ckel model (dashes) at four densities. The
      high-temperature relativistic correction is shown as a dotted
      line. To improve visibility on a log scale, the energies of the
      four isochores have been shifted by the N$_2$ molecule energy,
      $-$54.614969 Ha/atom, and multiplied by factors of 10, 100, 1,000,
      and 10,000 as indicated in the labels. The original energies are
      given in the Supplementary Material~\cite{SupMat}.}

  \label{TvsE}
\end{figure}

\begin{figure}[t]
  \begin{center}
        \includegraphics*[width=8.6cm]{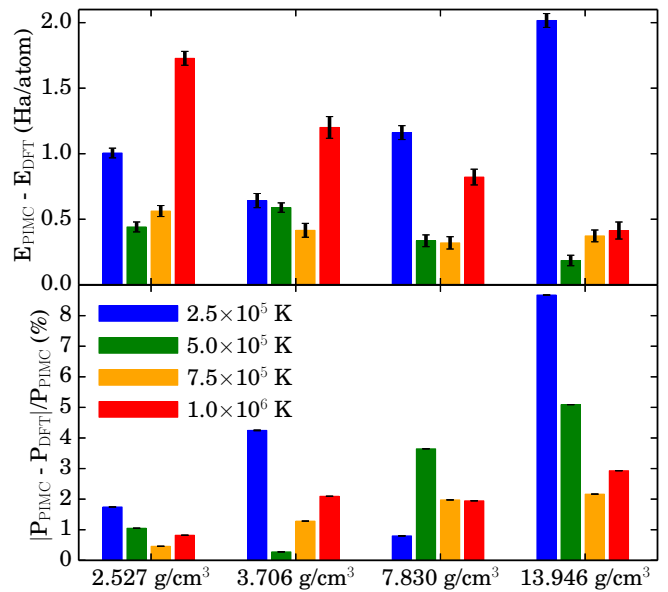}
  \end{center}

    \caption{Differences in PIMC and DFT-MD energies and pressures. The
      top panel shows energy differences, while the bottom panel shows
      the absolute relative error of pressure in per cent. One-$\sigma$
      errors in the differences are shown in black.}

  \label{EPerrors}
\end{figure}

DFT-MD~\cite{Marx2000} is an efficient, state-of-the-art,
first-principles method for zero and low temperatures (T $<$
1$\times$10$^6$ K). DFT formalism provides a mapping of the many-body
problem onto a single-particle problem with an approximate
exchange-correlation potential to describe many-body effects. In the
WDM regime, where temperatures are at or above the Fermi temperature,
the exchange-correlation functional is not explicitly designed to
accurately describe the electronic
excitations~\cite{Brown2013}. However, in our previous PIMC and DFT-MD
work~\cite{Driver2012}, we found existing DFT functionals to be
sufficiently accurate even at high temperatures.

DFT incorporates effects of finite electronic temperature by using a
Fermi-Dirac function to smear out the thermal occupation of
single-particle electronic states~\cite{Mermin1965}. As temperature
grows large, an increasing number of bands are required to account for
the occupation of excited states in the continuum, which typically
causes the efficiency of the algorithm to become intractable at
temperatures beyond 1$\times$10$^6$ K. In addition, pseudopotentials
replace the core electrons in each atom to improve efficiency. Here,
we are careful to avoid using DFT-MD at temperatures where the K shell electrons
undergo excitations and study those conditions with PIMC instead.

Progress has been made in orbital-free (OF) DFT and average-atom DFT
methods, which introduce additional approximations beyond standard
Kohn-Sham DFT-MD in order to improve the efficiency of the scaling
with temperature. OF-DFT approximates the free energy of the
homogeneous electron gas by a functional that is independent of the
single-particle orbitals~\cite{Lambert2006,Lambert2007}. The speed-up
gained has resulted in an significant trade-off in accuracy, but
recent OF-DFT developments have shown the method is potentially
capable of being competitive with
KS-DFT~\cite{Karasiev2013,Sjostrom2014}. In an effort to make even
greater gains in efficiency, DFT-based average-atom models make
further approximations based on solving for the electronic properties
of a single atom within the plasma~\cite{Rozsnyai2014}. Such models
have been shown to predict the electronic structure of the isolated
atoms well, and recent developments have begun a more consistent
treatment of many-body systems~\cite{Starrett2015}.  OF-DFT and
average-atom models are capable of simulating systems sizes up to a
few hundred particles, but ultimately, along with DFT-MD, they are
based on a ground-state framework, and it is important to develop more
accurate, finite-temperature methods with fewer approximations, such
as PIMC, to benchmark such calculations.

We employ standard Kohn-Sham DFT-MD simulation techniques for our
calculations of liquid and WDM matter states.  Simulations are
performed with the Vienna $Ab~initio$ Simulation Package
(VASP)~\cite{VASP} using the projector augmented-wave (PAW)
method~\cite{PAW}, and a NVT ensemble, regulated with a
Nos\'{e}-Hoover thermostat. Exchange-correlation effects are described
using the Perdew-Burke-Ernzerhof~\cite{PBE} generalized gradient
approximation. Electronic wave functions are expanded in a plane-wave
basis with a energy cut-off as high as 2000 eV in order to converge
total energy. For liquid simulations, we used 64-atom supercells with
a time-step of 1.5 fs. For WDM calculations, size convergence tests up
to a 24-atom simulation cell at temperatures of 10,000 K and above
indicate that total energies are converged to better than 0.1\% in a
24-atom simple cubic cell. We find, at temperatures above 250,000 K,
8-atom supercell results are sufficient since the kinetic energy far
outweighs the interaction energy at such high
temperatures~\cite{Driver2015Neon}. The number of bands in each
calculation is selected such that thermal occupation is converged to
better than 10$^{-4}$, which requires up to 8,000 bands in a 24-atom
cell at 1$\times$10$^6$ K. All simulations are performed at the
$\Gamma$-point of the Brillouin zone, which is sufficient for high
temperature fluids, converging total energy to better than 0.01\%
relative to a comparison with a converged grid of k-points.

\section{EOS OF LIQUID, WDM, AND PLASMA PHASES}

In this section, we report our DFT-MD and PIMC EOS results for the
liquid, WDM, and plasma regimes. The Supplementary
Material~\cite{SupMat} provides all of our computed pressure and
energy data. The VASP DFT-MD energies have been shifted by
$-$54.3064682071 Ha/atom in order to bring the PAW-PBE pseudopotential
energy in alignment with all-electron DFT calculations. The shift was
calculated by performing an all-electron atomic calculation with the
OPIUM code~\cite{OPIUM} and a corresponding isolated-atom calculation
in VASP.

In the liquid regime, we computed isochores with DFT-MD on a dense
grid of 15 densities spanning conditions from 1.5--3.7 g~cm$^{-3}$ and
10$^3$--5$\times$10$^4$~K, in order to accurately map out the
molecular dissociation transition. We extend the work of
Boates~\emph{et al.}~\cite{Boates2009} to higher temperatures and
lower pressures.  Our pair-correlation curves agree with the
experimental molecular bond length of 1.1~$\rm \AA$ at low temperature
and are generally consistent with the work of Boates~\emph{et al.} Our
dissociation curve was constructed by determining the temperature at
which the molecular lifetime reached 0.2 ps, which is the same cutoff
for molecular stability used by Boates and limits it to 15
vibrations. Consistent with previous
work~\cite{Radousky1986,Nellis1984,Nellis1991,Mazevet2001,Boates2009},
we find $(\partial P/ \partial T)_V<0$ in the dissociation region with
a first order dissociation transition at pressures near 78-90
GPa. Below 18 GPa, we find no $(\partial P/ \partial T)_V<0$ region
exists.

Fig.~\ref{fig:PTdiagram} shows the pressure-temperature phase diagram
with our dense grid of DFT-MD isochores in the liquid region, as well
as the dissociation and Hugoniot curves. The lower panel of
Fig.~\ref{fig:PTdiagram}, which includes a subset of our DFT-MD
isochores, shows the phase diagram ranging from solid to
low-temperature plasma phases. The solid phase boundaries, outlined
with solid lines, are reproduced from a variety of
experiments~\cite{Mills1986,Young1987,Olijnyk1990,Bini2000,Gregoryanz2002,Gregoryanz2007,Mukherjee2007,Goncharov2008}. The
melting curve is also reproduced from
experiments~\cite{Young1987,Mukherjee2007,Goncharov2008}, which agrees
with DFT-MD calculations~\cite{Donadio2010}, and displays a negative
slope with a triple point near 90 GPa and 1000 K.  We include phases
that have been predicted to be stable by a DFT random structure
searching algorithm at T=0~K~\cite{Pickard2009}, which have not been
seen by experiment.

The upper panel Fig.~\ref{fig:PTdiagram} is a magnified view of the
dissociation region, displaying a larger subset of the DFT-MD
isochores performed in our study. The molecules may dissociate into
polymeric or atomic fluid through a first order phase transition,
marked by the solid portion of dissociation line in the figure. As
pressure decreases, the dissociation curve reaches a critical point
near 78 GPa and 4100 K, marked by a white dot and a change to a dashed
line to indicate the transition is no longer first order. Starting at
18 GPa, where our DFT-MD data ends, we constructed a free energy
model~\cite{EbelingBook} with noninteracting atoms and molecules that
extends the dissociation curve to low pressures, marked by a thin,
dashed line. We postpone the discussion of the liquid Hugoniot until
Section VI.

In order to extend our nitrogen EOS into the WDM and plasma regimes,
we compute additional isochore data with DFT-MD and PIMC for
temperatures ranging up to $10^9$ K for four of the densities (2.5,
3.7, 7.8, and 13.9~g$\,$cm$^{-3}$).  Fig.~\ref{TvsP} shows the data
computed for the four isochores and compares pressures obtained for
nitrogen from PIMC and DFT-MD.  Likewise, Fig.~\ref{TvsE} compares
internal energies and also compares with results from the
Debye-H\"{u}ckel model~\cite{DebyeHuckel}, constructed for the fully
ionized plasma. Using a relativistic, fully-ionized
model~\cite{Landau1980}, we also show the magnitude of the
relativistic correction to the internal energy, which results in a
14\% change at the high-temperature limit. There is not a significant
relativistic correction to the pressure. In both pressure and energy,
we find good agreement between PIMC and DFT-MD results in the
temperature range of $5.0\times10^5-1\times10^6$ K. At a temperature
of $2.5\times10^5$ K, the PIMC free-particle nodes start to become
insufficient for describing bound electronic states, and the results
begin to deviate significantly from that of DFT-MD. At high
temperature, the PIMC pressures and energies converge to the weakly
interacting plasma limit, in agreement with the classical
Debye-H\"{u}ckel model.

Fig.~\ref{EPerrors} shows the differences between the PIMC and DFT-MD
pressures and energies as a function of temperature in the overlap
regime where both methods operate efficiently. DFT-MD and PIMC
internal energies differ by at most 2 Ha/atom and pressures differ by
less than 8\% in the temperature range of $2.5\times10^5-1\times10^6$
K. The size of the discrepancy between our PIMC and DFT-MD results
also places an approximate limit on the magnitude of the correction
that a new free-energy functional, such as those used in OF-DFT, can
change existing KS-DFT results. Typically, the error is largest at the
lowest and highest temperatures.  This is possibly because, at low
temperature, the PIMC free-particle nodes are expected to breakdown,
while, at high temperature, the DFT exchange-correlation functional
and pseudopotential may breakdown. The pseudopotential, with a frozen
1s core, may also begin to leave out excitation effects at
temperatures close to 10$^6$ K.  In our previous studies, we found it
is not uncommon for one third of the energy discrepancy at $10^6$ K to
be attributed to pseudopotential
error~\cite{Driver2012,Driver2015Neon,Driver2015Oxygen}.

Together, Figs.~\ref{TvsP} and ~\ref{TvsE} show that the DFT-MD and
PIMC methods form a coherent EOS over all temperatures ranging from
condensed matter to the WDM and plasma regimes. The good agreement
between PIMC and DFT-MD indicates that DFT exchange-correlation
potential remains valid even at high temperatures and that the PIMC
free-particle nodal approximation is valid as long as the 2s state is
sufficiently ionized. The analytic Debye-H\"{u}ckel models agree well
with PIMC at high temperatures, but the Debye-H\"{u}ckel model does
not include bound states and, therefore, cannot describe low
temperatures.

\begin{figure}[t]
  \begin{center}
        \includegraphics*[width=8.6cm]{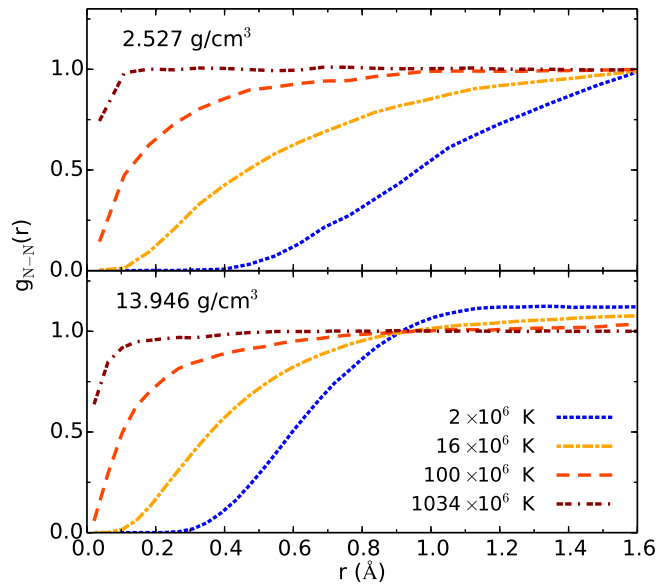}
  \end{center}

    \caption{Nuclear pair-correlation functions for nitrogen from PIMC
      over a wide range of temperatures and densities.}

  \label{fig:gofrNN}
\end{figure}

\begin{figure}[t]
  \begin{center}
        \includegraphics*[width=8.6cm]{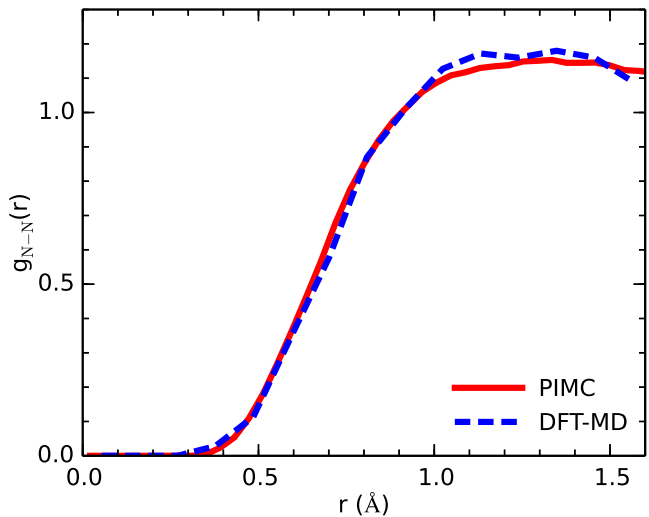}
  \end{center}

    \caption{Comparison of PIMC and DFT nuclear pair-correlation
      functions for nitrogen at a temperature of 1$\times$10$^6$ K and
      a density of 13.946 g$\,$cm$^{-3}$.}

  \label{fig:gofrNN1millionK}
\end{figure}

\section{PAIR-CORRELATION FUNCTIONS}

In this section, we study pair-correlation
functions~\cite{Militzer2009JPAM} in order to understand the evolution
of the fluid structure and ionization in nitrogen plasmas as a
function of temperature and density.

Fig.~\ref{fig:gofrNN} shows the nuclear pair-correlation functions,
$g(r)$, computed with PIMC over a temperature range of
$2\times10^6-1.034\times10^{9}$ K and for densities of 2.527 and
13.946 g$\,$cm$^{-3}$. Atoms are kept farthest apart at low
temperatures due to a combination of Pauli exclusion among bound
electrons and Coulomb repulsion. As temperature increases, kinetic
energy of the nuclei increases, making it more likely to find atoms at
close range. In addition, the atoms become increasingly ionized, which
gradually reduces the Pauli repulsion, but increases the ionic Coulomb
repulsion. As density increases, the likelihood of finding two nuclei
at close range slightly rises. At high temperatures, the system
approaches the Debye-H\"{u}ckel limit, behaving like a weakly
correlated system of screened Coulomb charges.

Fig.~\ref{fig:gofrNN1millionK} compares the nuclear
pair-correlation functions of PIMC and DFT-MD at a temperature of
1$\times$10$^6$ K in an 8-atom cell at a density of 13.946
g$\,$cm$^{-3}$. The overlapping $g(r)$ curves verify that PIMC and DFT
predict consistent structural properties.

\begin{figure}
  \begin{center}
        \includegraphics*[width=8.6cm]{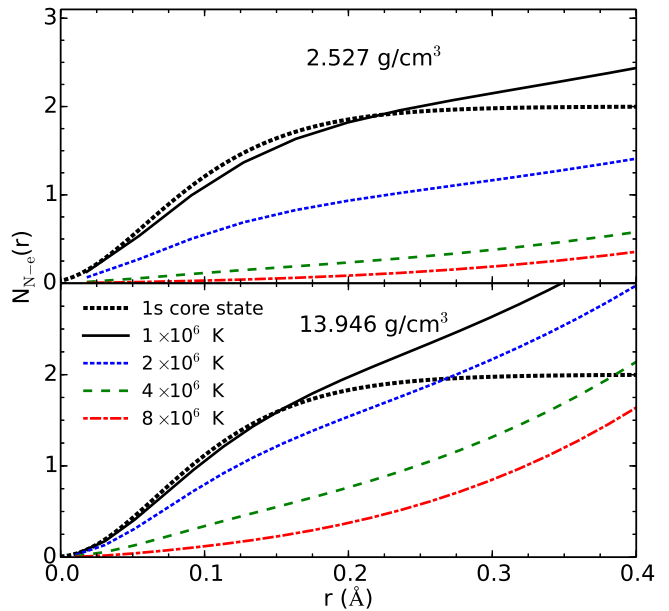}
  \end{center}

    \caption{N(r) function representing the number of electrons
      contained in a sphere of radius, $r$, around an nitrogen
      nucleus. PIMC data at four temperatures is compared with the
      analytic 1s core state.}

  \label{fig:nr}
\end{figure}

Fig.~\ref{fig:nr} shows the integral of the nucleus-electron pair correlation
function, $N_{N-e}(r)$, which represents the average number of electrons
within a sphere of radius $r$ around a given nucleus. At the lowest
temperature, 1$\times$10$^6$ K, we find that the 1s core state is
always fully occupied, as it agrees closely with the result of an
isolated 1s state. As temperature increases, the atoms are gradually
ionized and electrons become unbound, causing $N_{N-e}(r)$ to decrease. At
higher density, an even higher temperature is required to fully ionize
the atoms, indicating that the 1s ionization fraction decreases with
density.

There are two important physical points to note about this
result. First, it is clear that 1s ionization fraction is not affected
by pressure ionization in the considered density range, which is
supported by the fact that the nuclei are not yet close enough for
Pauli exclusion to trigger the ionization of the 1s state. Pauli
exclusion effects decay on the scale of $\sim 0.04$ $\rm \AA$ (size of
1s orbital), while Fig.~\ref{fig:gofrNN1millionK} shows that the
nuclei remain at least 0.3 $\rm \AA$ apart at our highest
density. Secondly, we note that in our work on dense
oxygen~\cite{Driver2015Oxygen} we performed all-electron DFT-MD
calculations and found that the 1s ionization fraction for a fixed
temperature decreases because the Fermi energy shifts to higher
energies more rapidly than the 1s state shifts towards the continuum
when density increases. Thus, the decrease in the 1s ionization
fraction in Fig.~\ref{fig:nr} at a fixed temperature with increasing
density is due to a rapid shift of the Fermi energy. Eventually, the
1s ionization fraction will increase when density is high enough to
push the 1s states into the continuum, but we have not studied such
densities here.

\begin{figure}
  \begin{center}
        \includegraphics*[width=8.6cm]{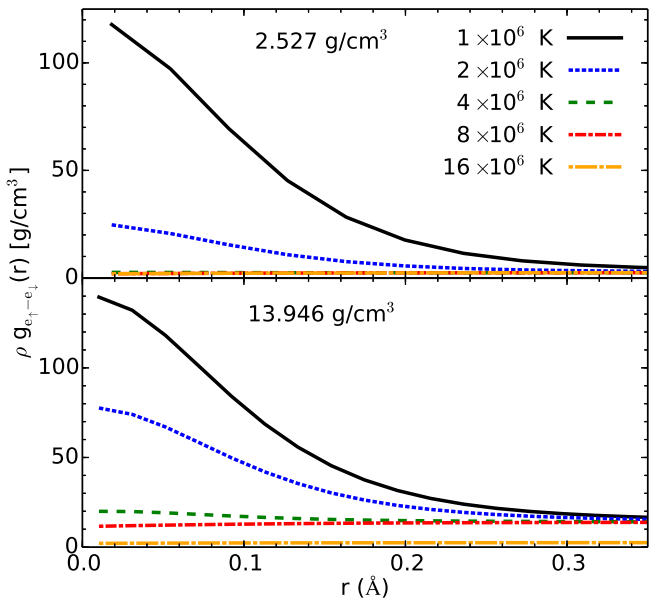}
  \end{center}

    \caption{The electron-electron pair-correlation functions for
      electrons with opposite spins computed with PIMC.}

  \label{fig:gofreeopp}
\end{figure}

\begin{figure}
  \begin{center}
        \includegraphics*[width=8.6cm]{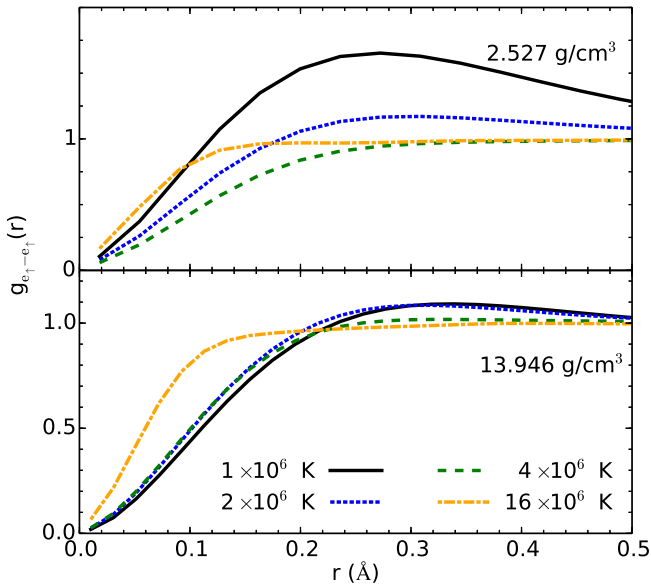}
  \end{center}

    \caption{The electron-electron pair-correlation functions for
      electrons with parallel spins computed with PIMC.}

  \label{fig:gofreepara}
\end{figure}

Fig.~\ref{fig:gofreeopp} shows electron-electron pair correlations for
electrons having opposite spins. The function is multiplied by the
density $\rho$, so that the integral under
the curves is proportional to the number of electrons. The electrons are
most highly correlated for low temperatures, which reflects that
multiple electrons occupy bound states around a given nucleus. As
temperature increases, electrons are thermally excited, decreasing the
correlation among each other. The positive correlation at short
distances increases with density, consistent with a lower ionization
fraction.

Fig.~\ref{fig:gofreepara} shows electron-electron pair correlations
for electrons with parallel spins.  The positive correlation at
intermediate distances reflects that different electrons with parallel
spins are bound to a given nucleus. For short separations, electrons
strongly repel due to Pauli exclusion and the functions decay to zero.
As density increases, the peak at intermediate distances decreases,
which clearly shows the effect of pressure ionization on the L
shell. These orbitals are much larger than the 1s state and are
therefore subject to Pauli exchange with nearby nuclei. As temperature
increases, electrons become less bound, which also causes the
correlation to become more like an ideal fluid.

\section{ELECTRONIC DENSITY OF STATES}

In this section, we report DFT-MD results for the electronic density
of states (DOS) as a function of temperature and density in order to
gain further insight into temperature and pressure ionization effects.

\begin{figure}
  \begin{center}
        \includegraphics*[width=8.6cm]{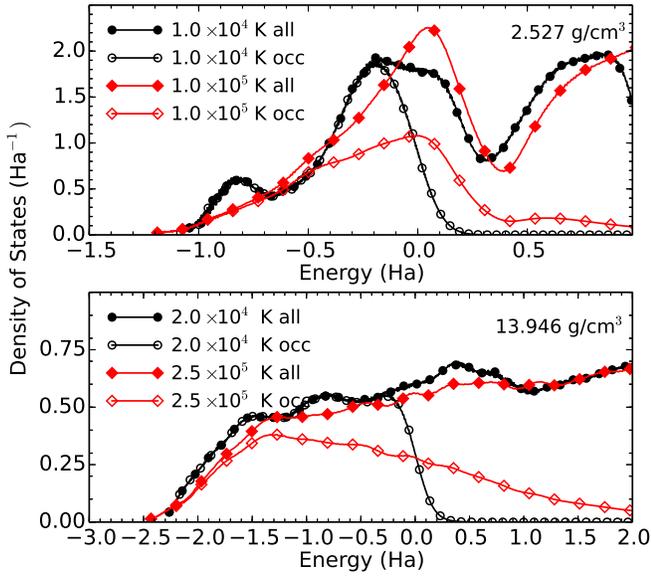}
  \end{center}

  \caption{Temperature dependence of the total (all) and occupied
    (occ) electronic DOS of dense, fluid nitrogen at densities of 2.53
    and 13.95 g$\,$cm$^{-3}$. Each DOS curve has had the relevant
    Fermi energy for each temperature subtracted from it.}

  \label{fig:DOS}
\end{figure}

Fig.~\ref{fig:DOS} shows the total and occupied DOS at two
temperatures and two densities. Results were obtained by averaging
over ten uncorrelated snapshots chosen from a DFT-MD
trajectory. Smooth curves were obtained by using a 4$\times$4$\times$4
k-point grid and applying a Gaussian smearing of 2~eV. The eigenvalues
of each snapshot were shifted so that the Fermi energies align at
zero. The integral of the DOS is normalized to 1.

At low temperature and density, the general structure is composed of
two peaks below the Fermi energy, representing the atomic 2s and 2p
states. The peaks broaden and merge at higher temperatures and
densities as they become ionized. For higher density, the total DOS
resembles that of an ideal plasma. For lower densities, a dip in the
DOS indicates beginning of the continuous spectrum of conducting
states.
At the lowest temperature ($\sim$10$^4$ K) shown for each
density, the majority of occupied states lie below the Fermi energy.
At the higher temperature ($\sim$10$^5$ K), a significant fraction of the
occupied states now lie above the Fermi energy as the second shell
becomes ionized. 
Finally, we note that the Fermi energy plays the role of the chemical
potential in the Fermi-Dirac distribution, which shifts towards more
negative values as the temperature is increased.  Because we subtract
the Fermi energy from the eigenvalues, the peak shifts to higher
energies with increasing temperature. The fact that the peaks are
embedded into a dense, continuous spectrum of eigenvalues indicates
that they are conducting states.

\section{SHOCK COMPRESSION}

\begin{figure}
  \begin{center}
        \includegraphics*[width=8.6cm]{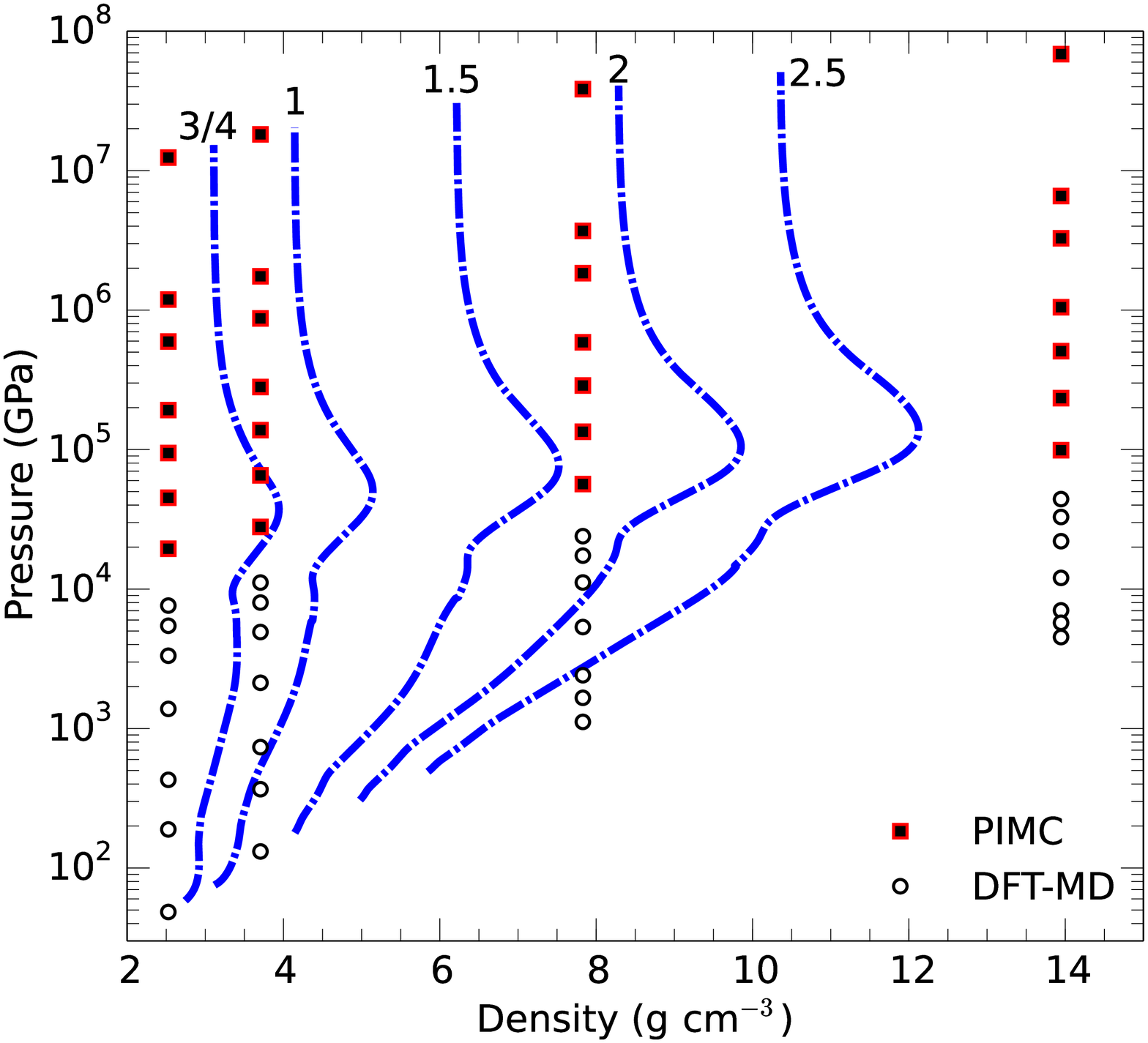}
  \end{center}

  \caption{Shock Hugoniot curves for different initial densities
    ranging from 0.75- to 2.5-fold the density of solid N$_2$, 1.035
    g$\,$cm$^{-3}$, at ambient pressure.}

  \label{fig:hugoniot1}
\end{figure}

\begin{figure}
  \begin{center}
        \includegraphics*[width=8.6cm]{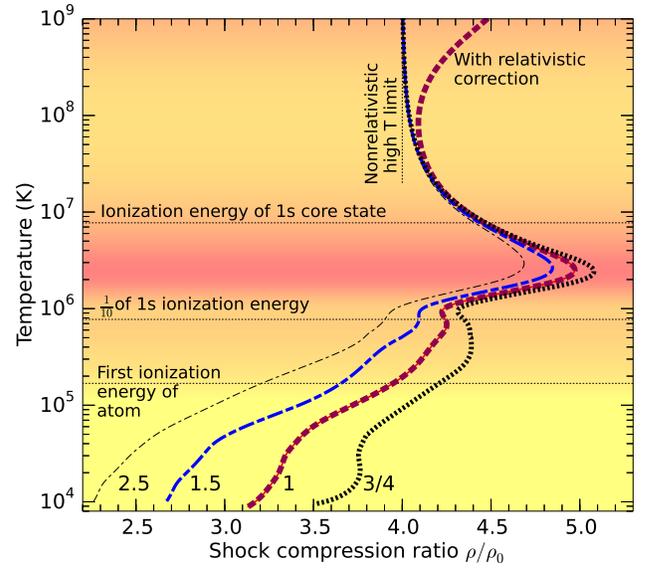}
  \end{center}

    \caption{Hugoniot curves as a function of the shock compression
      ratio for different initial densities as plotted in
      Fig.~\ref{fig:hugoniot1}. The 1-fold curve is shown with (dashed
      line) and without (solid line) the relativistic correction. The
      dark shaded marks the temperature range of highest compression.}

  \label{fig:hugoniot2}
\end{figure}

\begin{figure}
  \begin{center}
        \includegraphics*[width=8.6cm]{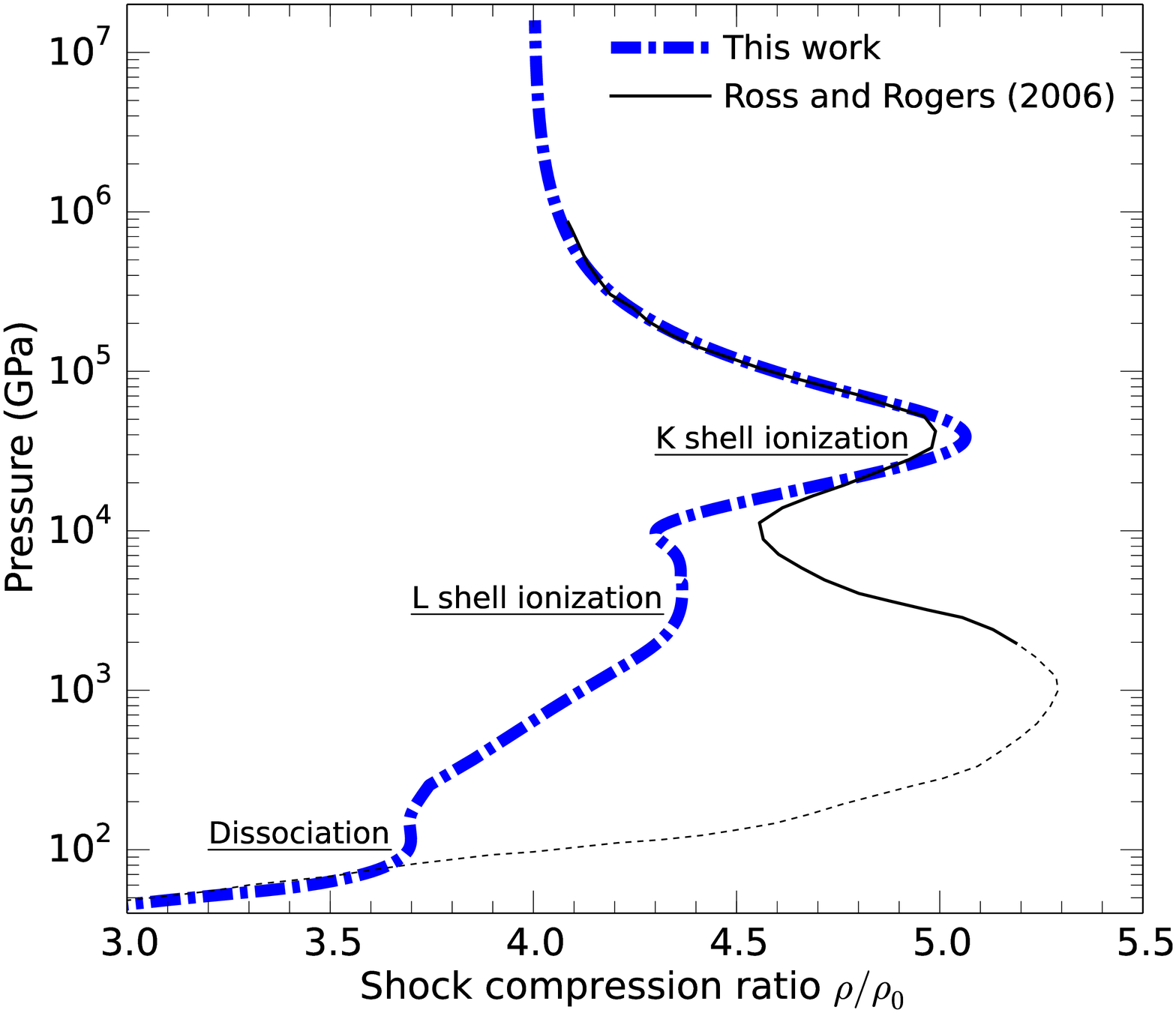}
  \end{center}

  \caption{Comparison of our combined PIMC and DFT-MD Hugoniot curve
    with predictions of ACTEX plasma model calculations by Ross and
    Rogers~\cite{Ross2006}. The dashed line portion of the plasma
    model curve indicates where the ACTEX results were interpolated to
    match experimental data below 100 GPa. The initial density was
    $\rho_0=0.8076$ g$\,$cm$^{-3}$ ($V_0$= 28.80 \AA$^3$/atom).}

  \label{fig:NHugCompACTEX}
\end{figure}

Dynamic shock compression experiments allow one to measure the EOS and
other physical properties of hot, dense fluids.  Such experiments are
often used to determine the principal Hugoniot curve, which is the
locus of final states that can be obtained from different shock
velocities. A number of Hugoniot measurements have been made for
nitrogen~\cite{Dick1970,Nellis1980,Nellis1984,Radousky1986,Moore1989,Nellis1991,Chau2003}.
Density functional theory has been validated by experiments as an
accurate tool for predicting the shock compression of a variety of
different materials~\cite{Root2010,Mattsson2014}, including
nitrogen~\cite{Kress2000,Mazevet2001}.

In the course of a shock wave experiment, a material whose initial
state is characterized by an internal energy, pressure, and volume
($E_0,P_0,V_0$) will change to a final state denoted by ($E,P,V$)
while conserving mass, momentum, and energy. This leads to the
Rankine-Hugoniot relation~\cite{Ze66},
\begin{equation}
(E-E_0) + \frac{1}{2} (P+P_0)(V-V_0) = 0.
\label{hug}
\end{equation}

Here, we compute the Hugoniots from the first-principles EOS data
reported in the Supplementary material~\cite{SupMat}. The pressure and
internal energy data points were interpolated with bi-cubic spline
functions in $\rho-T$ space. For the initial state, we used the energy
of an isolated (P$_0$ = 0) nitrogen molecule, E$_0$ = $-$109.2299
Ha/N$_2$.  V$_0$ was determined by the density, $\rho_0=1.035$
g$\,$cm$^{-3}$, of solid nitrogen in the $Pa\bar{3}$
phase~\cite{Venables1973}. The resulting Hugoniot curve has been
plotted in $T$-$P$ and $P$-$\rho$ spaces in Figs.~\ref{TvsP}
and~\ref{fig:hugoniot1}, respectively.

Samples in shock wave experiments may be pre-compressed inside of a
diamond anvil cell in order to reach much higher final densities than
are possible with a sample at ambient conditions. This technique
allows shock wave experiments to probe density-temperature consistent
with planetary and stellar interiors~\cite{MH08}. Therefore, we
repeated our Hugoniot calculation starting with initial densities
ranging from a 0.75 to a 2.5-fold of the density typically used in
shock-compression experiments (0.808 g$\,$cm$^{-3}$).
Fig.~\ref{fig:hugoniot1} shows the resulting family of Hugoniot
curves. While starting from an initial density of 0.808 g$\,$cm$^{-3}$
leads to a maximum shock density of 5.15 g$\,$cm$^{-3}$ (4.97-fold
compression), a 2.5-fold pre-compression yields a much higher maximum
shock density of 12.13 g$\,$cm$^{-3}$ (4.69-fold compression).
Alternatively, such extreme densities can be reached with double and
triple shock experiments.

Fig.~\ref{fig:hugoniot2} shows the temperature dependence of the
shock-compression ratio for the four representative Hugoniot curves
from Fig.~\ref{fig:hugoniot1}. In the high-temperature limit, all
curves converge to a compression ratio of 4, which is the value of a
nonrelativistic, ideal gas. We also show the magnitude of the
relativistic correction to the Hugoniot in the high-temperature
limit. The shock compression and structure along the Hugoniot is
determined by the excitation of internal degrees of freedom, such as
dissociation and ionization processes, which increases the
compression, and, in addition, the interaction effects, which decrease
the compression~\cite{Mi06}. Consistent with our studies of other
elements, we find that an increase in the initial density leads to a
slight reduction in the shock compression ratio
(Fig.~\ref{fig:hugoniot2}) because particles interact more strongly at
higher density.

\begin{figure}
  \begin{center}
        \includegraphics*[width=8.6cm]{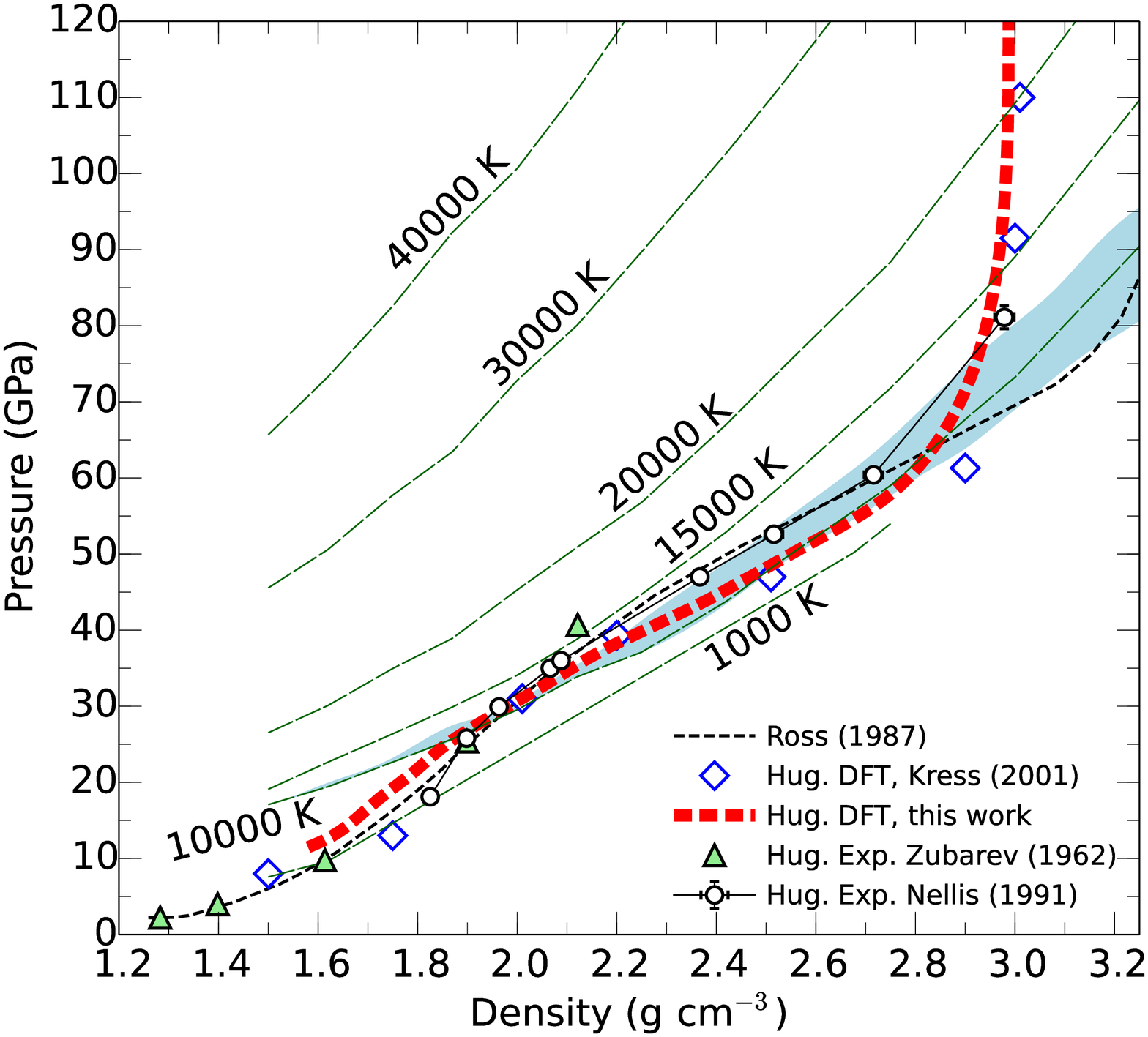}
  \end{center}

\caption{Comparison of the liquid DFT-MD Hugoniot with the experiments
  of Nellis~\emph{et al.}~\cite{Nellis1991} and Zubarev~\emph{et
    al.}~\cite{Zubarev1962} and the theory of Ross~\emph{et
    al.}~\cite{Ross1987} (variational fluid theory) and Kress~\emph{et
    al.}~\cite{Kress2000} (DFT-MD). The blue shaded region indicates
  the region of dissociation with $(\partial P/ \partial T)_v<0$ in
  the phase diagram of Fig.~\ref{fig:PTdiagram}. Our Hugoniot passes
  through this region, but there is no evidence of cooling along the
  principal Hugoniot curve. The green dashed lines show isotherms from
  our DFT-MD simulations.}

  \label{fig:hugoniotLowT}
\end{figure}

\begin{figure}
  \begin{center}
        \includegraphics*[width=8.6cm]{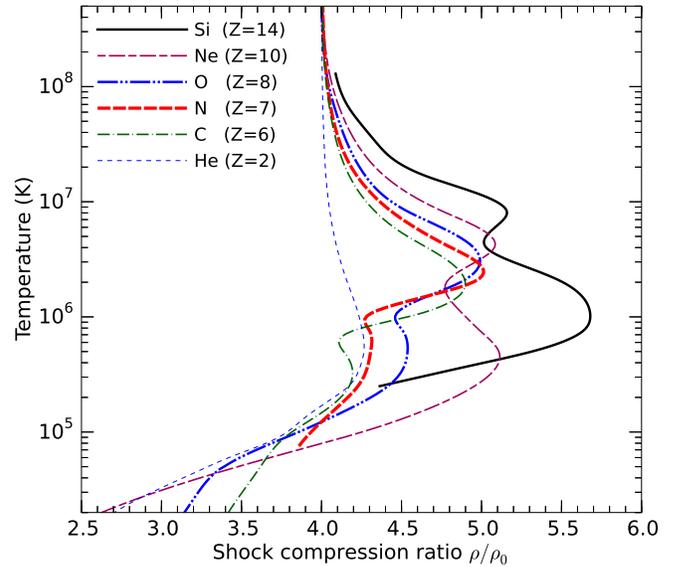}
  \end{center}

\caption{Comparison of the shock Hugoniot curves for different
  materials. The initial volume $V_0$ has been chosen such that
  the density of the electrons is the same for all materials
  ($V/N_e=3.586$~\AA$^3$). The various maxima in compression
  corresponds to excitations of electrons in the first and second
  electron shells.}

  \label{fig:AllHugComparison}
\end{figure}

\begin{figure}
  \begin{center}
        \includegraphics*[width=8.6cm]{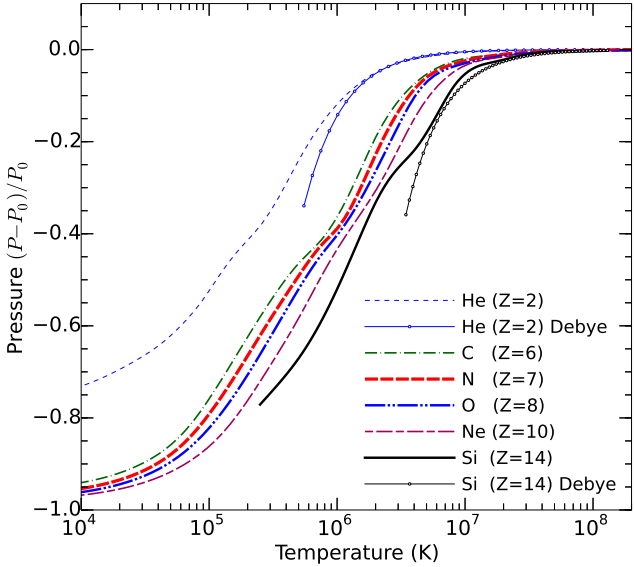}
  \end{center}

\caption{Pressure vs. temperature is shown for isochores of different
  materials. The pressure of a fully-ionized, non-interacting plasma,
  $P_0$, has been removed in order to compare the excess pressure due
  to interactions. The densities have been chosen such that electronic
  density is the same for all materials ($V/N_e=0.8966$~\AA$^3$). This
  electronic density corresponds to the high-temperature limit of
  4-fold compression of the shock Hugoniot curves in
  Fig.~\ref{fig:AllHugComparison}. The Debye model has been included for
  helium and silicon.}

  \label{fig:PComparison}
\end{figure}

\begin{figure}
  \begin{center}
        \includegraphics*[width=8.6cm]{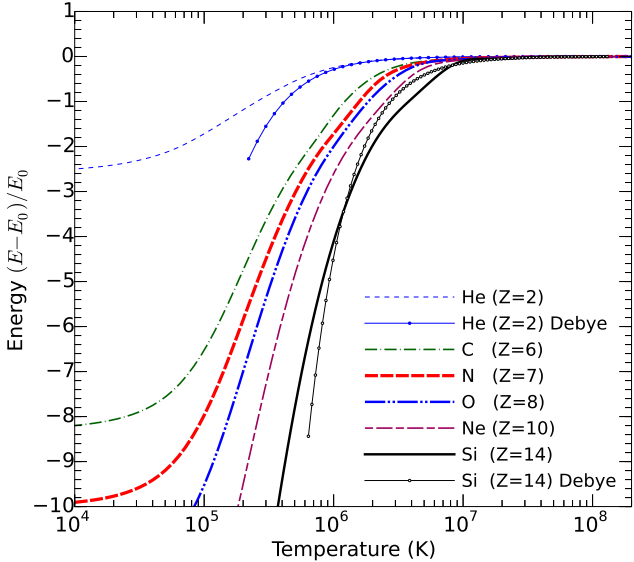}
  \end{center}

\caption{ Internal energy vs. temperature is shown for the isochores
  in Fig.~\ref{fig:PComparison}. The energy contribution from a
  fully-ionized, non-interacting plasma, $E_0$, has been removed in
  order to compare only the interaction effects.}

  \label{fig:IEComparison}
\end{figure}

For the lowest two initial densities, the shock compression ratio in
Fig.~\ref{fig:hugoniot2} exhibits two maxima as a function of
temperature, which can be attributed to the ionization of electrons in
the K (1s) and L (2s+2p) shells. On the principal Hugoniot curve, the
first maximum of $\rho/\rho_0$=4.26 occurs at temperature of
$6.77\times 10^5\,$ K (58.3 eV) , which is well above the first and
second ionization energies of the nitrogen atom, 14.53 and 29.60 eV. A
second compression maximum of $\rho/\rho_0$=4.97 is found for a
temperature of $2.55\times 10^6\,$ K (220 eV), which can be attributed
to a substantial ionization of the 1s core states. For an isolated
nitrogen atom, the 1s ionization energy is 667.05 eV. However,
fractional ionization is expected to occur at much lower temperature
already.  This is consistent with the ionization process we observe in
Fig.~\ref{fig:nr}, where the charge density around the nuclei is
reduced over the range of $2-8\times10^6$ K. Since DFT-MD simulations,
which use pseudopotentials to replace core electrons, cannot access
the regime of core ionization, both PIMC and DFT-MD are needed to
determine all features along the principal Hugoniot curve.

Fig.~\ref{fig:NHugCompACTEX} compares our combined PIMC and DFT-MD
Hugoniot curve with predictions from the ACTEX calculations by Ross
and Rogers~\cite{Ross2006}. We find very good agreement for P $\ge$
20,000 GPa, which includes a compression peak due to the ionization of
K shell and confirms the strengths of the ACTEX method in highly
ionized regimes with weak-to-moderate Coulomb coupling.  While the
K shell peak pressures agree almost perfectly in pressure, the ACTEX
predicts a maximum compression ratio that is 0.07 lower than predicted
by our PIMC simulations.  In the pressure range from 2,000 to 20,000
GPa, where ionization of the L shell occurs, we find that the ACTEX
model substantially overestimates the shock compression. In the range
of 100 to 2000 GPa (dashed line in Fig.~\ref{fig:NHugCompACTEX}), Ross
and Rogers interpolated their Hugoniot curve based on a collection of
previous ACTEX calculations for other light
elements~\cite{Rozsnyai2001} and available experimental data below 100
GPa~\cite{Nellis1991}. Therefore, it is not too surprising that PIMC
and the analytic model disagree by up to 20\% in the pressure.  The
comparison shows the importance of using first-principles methods such
as PIMC and DFT-MD to correctly predict the ionization compression
peaks of the Hugoniot curve in more strongly coupled regimes. With
DFT-MD, we are also able to capture the sharp change in slope in the
Hugoniot curve, which is associated with dissociation as internal
energy is absorbed to break the molecular bond.

Fig.~\ref{fig:hugoniotLowT} shows a magnified view of the low-pressure
Hugoniot in the dissociation region. Our DFT-MD Hugoniot generally
agrees well with the experimental data of Nellis~\emph{et
  al.}~\cite{Nellis1991} and previous DFT-MD
calculations~\cite{Kress2000}. DFT-MD accurately captures the sharp
increase in compressibility in the dissociation transition region,
while the Ross model underestimates the compressibility more or less
depending on the parameterization~\cite{Ross1987}. Slight deviations
with experiment tend to lie near the region of $(\partial P/ \partial
T)_V<0$, marked by the blue shaded region. The discrepancy could
either be due to impedance matching difficulties in experiment or
shortcomings of DFT-MD approximations. A negative $(\partial P/
\partial T)_V$ region and molecular dissociation can, in principle,
trigger a shock wave to split into two separate
waves~\cite{ZeldovichBook}. This occurs when the shock speed is not
monotonously increasing with particle speed. However, this is not
predicted to occur based on our DFT-MD EOS, and we find it unlikely
that this hypothesis can explain the discrepancy between the
theoretical and experimental results in
Fig. ~\ref{fig:hugoniotLowT}. We also note that including zero point
motion has a negligible affect on the Hugoniot curve.

\section{EOS COMPARISON OF FIRST- AND SECOND-ROW PLASMAS}

Using PIMC and DFT-MD, we have computed the first principles EOS and
shock Hugoniot curves for several materials in the the WDM and dense,
plasma regime. In this section, we compare our collective sets of data
and discuss some of the trends we have observed.

Fig.~\ref{fig:AllHugComparison} compares our computed shock Hugoniot
curves from simulations of He~\cite{Mi09}, C~\cite{Driver2012},
O~\cite{Driver2015Oxygen}, Ne~\cite{Driver2015Neon}, and
Si~\cite{Militzer2015Silicon} in the WDM and plasma regimes. The
Hugoniot curve comparison shows distinct compression maxima for all
materials, but the maxima and structure along the Hugoniot depend
strongly on the atomic number, Z, which is directly connected to
internal degrees of freedom and interaction effects~\cite{Mi06}. We
find the shock Hugoniot compression maxima, corresponding to K and
L shell ionization, increase in both compression and temperature with
the atomic number, Z. This is not unexpected because the binding
energy scales as Z$^2$, which means a higher temperature is needed to
reach the regime of ionization. When this happens, a larger energy
difference, $E-E_0$, must be compensated by the $P(V-V_0)$ term in
Eq.~\ref{hug}. Even though the pressure increases with ionization
also, we still see a higher shock compression for higher Z materials
in Fig.~\ref{fig:AllHugComparison}.

Figs.~\ref{fig:PComparison} and \ref{fig:IEComparison} compare the
pressure and internal energies of the same set of materials in the
Hugoniot curve comparison. The plots compare the excess pressure and
energy, where the ideal Fermi gas contributions have been removed in
order to compare only interaction effects, which become important for
$T<10^8$~K when electrons start to occupy the K shell. For higher Z,
this occurs at higher temperature, which explains the trends seen in
Figs. 15-17. The Debye model can capture only the high temperature
limit of this trend since it cannot describe the occupation of the K
shell. There is a visible softening of the slope in the pressure and
internal energy curves for temperatures around 10$^6$~K, which
corresponds to the intermediate regime between K and L shell
ionization. As expected, the onset of the slope-softening occurs at
higher temperatures for higher Z elements.

We note that, for each material, that we have computed consistent,
overlapping results with both DFT-MD and PIMC at temperatures near
10$^6$~K. The agreement implies that our zero-temperature, DFT
exchange-correlation potential (PBE) remains valid for a large set of
materials at high temperatures and that the free-particle nodal
approximation is accurate in PIMC when the K shell electrons are bound
and L-shell is partially ionized.


\section{CONCLUSIONS}

In this work, we have used DFT-MD and PIMC to compute liquid and WDM
states of nitrogen to provide an EOS witch bridges the condensed
matter and warm dense matter regimes. In the liquid regime, we have
extended the phase diagram beyond previous studies by computing the
dissociation curve for a broader region of conditions and extending
the Hugoniot to the WDM regime. In the WDM regime, we have combined
PIMC with DFT-MD to construct a coherent EOS for nitrogen over a wide
range of densities and temperatures. The two methods produce
consistent pressures and energies in temperature range of
5.0$\times$10$^5$--1$\times$10$^6$ K.  At high temperatures, our EOS
converges to the analytic Debye-H\"{u}ckel result for weakly
interacting plasmas.
Nuclear and electronic pair-correlations reveal a temperature- and
pressure-driven ionization process, where temperature-ionization of
the 1s state is suppressed, while other states are efficiently ionized
as temperature and density increases. Temperature-density dependence
of the electronic density of states confirms the temperature- and
pressure-ionization behavior observed in the pair-correlation data.
Lastly, we find the ionization imprints a signature on the shock
Hugoniot curves and that PIMC simulations are necessary to determine
the state of the highest shock compression. By combining our liquid
DFT-MD data with our WDM data, we provide a first-principles Hugoniot
that matches experiment at low pressures and extends to the classical
plasma regime. Our Hugoniot and equation of state will help to build
more accurate models for astrophysical applications and energy
applications.

\begin{acknowledgments}
  
  This research is supported by the U. S. Department of Energy, grant
  DE-SC0010517. Computational support was provided by NERSC and the
  Janus supercomputer, which is supported by the National Science
  Foundation (Grant No. CNS-0821794), the University of Colorado, and
  the National Center for Atmospheric Research.

\end{acknowledgments}


\end{document}


\title{First-Principles Equation of State Calculations of Warm Dense Nitrogen}

\author{K. P. Driver}

\email{kdriver@berkeley.edu}

\homepage{http://militzer.berkeley.edu/~driver/}

\affiliation{Department of Earth and Planetary Science, University of California, Berkeley, California 94720, USA}

\author{B. Militzer}

\affiliation{Department of Earth and Planetary Science, University of California, Berkeley, California 94720, USA}

\affiliation{Department of Astronomy, University of California, Berkeley, California 94720, USA}

\date{\today}

\maketitle
 
\begin{longtable}[c]{lrrr}
\caption{Nitrogen EOS table of pressures and internal energies at
  density-temperature conditions simulated in this work. Relativistic
  corrections are included for $T \geq 8\times10^6$~K. The numbers in
  parentheses indicate one-sigma statistical uncertainties of the
  DFT-MD and PIMC simulations.\label{long}}\\

\hline
\hline\\[-6pt]
 Method  & T (K) & P (GPa)  &  E (Ha/atom)\\[1pt]
\hline\\[-6pt]
\endfirsthead

\hline\\[-6pt]
\multicolumn{4}{c}{Table \ref{long} \emph{Continued}.}\\[1pt]
\hline\\[-6pt]
Method  & T (K) & P (GPa)  &  E (Ha/atom)\\[1pt]
\hline\\[-8pt]
\endhead

\hline
\endfoot

\hline
\hline
\endlastfoot

$\rho=1.50000$ g$\,$cm$^{-3}$ & & &\\[1pt]
\hline\\[-6pt]
DFT-MD & 1000  & 7.56(7)  & $-54.6030(1)$\\
DFT-MD & 2000  & 9.71(8)  & $-54.5953(1)$\\
DFT-MD & 2500  & 10.49(8) & $-54.5917(1)$\\
DFT-MD & 3500  & 12.3(1)  & $-54.5843(2)$\\
DFT-MD & 5000  & 14.3(2)  & $-54.5741(3)$\\
DFT-MD & 6000  & 15.3(1)  & $-54.5663(3)$\\
DFT-MD & 7500  & 16.3(3)  & $-54.5502(8)$\\
DFT-MD & 10000 & 17.1(1)  & $-54.511(1) $\\
DFT-MD & 15000 & 19.1(2)  & $-54.422(2) $\\
DFT-MD & 20000 & 26.5(3)  & $-54.349(1) $\\
DFT-MD & 30000 & 45.5(5)  & $-54.226(1) $\\
DFT-MD & 40000 & 65.7(4)  & $-54.0940(9)$\\
DFT-MD & 50000 & 88.1(7)  & $-53.9448(9)$\\
\hline\\[-6pt]
$\rho=1.62000$ g$\,$cm$^{-3}$ & & &\\[1pt]
\hline\\[-6pt]
DFT-MD & 1000  & 9.64(5)  & $-54.6013(1)$\\
DFT-MD & 2000  & 11.85(7) & $-54.5932(1)$\\
DFT-MD & 2500  & 12.66(5) & $-54.5892(8)$\\
DFT-MD & 3500  & 14.5(1)  & $-54.5820(1)$\\
DFT-MD & 5000  & 16.9(1)  & $-54.5699(2)$\\
DFT-MD & 6000  & 18.2(1)  & $-54.5616(1)$\\
DFT-MD & 7500  & 19.8(2)  & $-54.5488(3)$\\
DFT-MD & 10000 & 19.4(2)  & $-54.503(1) $\\
DFT-MD & 15000 & 22.7(3)  & $-54.422(1) $\\
DFT-MD & 20000 & 30.1(3)  & $-54.350(1) $\\
DFT-MD & 30000 & 50.6(4)  & $-54.2267(5)$\\
DFT-MD & 40000 & 73.3(3)  & $-54.0987(8)$\\
DFT-MD & 50000 & 98.7(5)  & $-53.9528(6)$\\
\hline\\[-6pt]
$\rho=1.75000$ g$\,$cm$^{-3}$ & & &\\[1pt]
\hline\\[-6pt]
DFT-MD & 2000  & 15.39(9) & $-54.5911(1)$\\
DFT-MD & 2500  & 16.47(7) & $-54.5872(1)$\\
DFT-MD & 3500  & 18.6(1)  & $-54.5791(1)$\\
DFT-MD & 5000  & 20.6(1)  & $-54.5679(2)$\\
DFT-MD & 6000  & 22.4(1)  & $-54.5595(2)$\\
DFT-MD & 7500  & 23.2(3)  & $-54.5418(7)$\\
DFT-MD & 10000 & 22.7(2)  & $-54.498(1) $\\
DFT-MD & 15000 & 26.4(5)  & $-54.420(2) $\\
DFT-MD & 20000 & 35.0(4)  & $-54.3539(9)$\\
DFT-MD & 30000 & 57.7(6)  & $-54.231(1) $\\
DFT-MD & 40000 & 82.6(6)  & $-54.1030(7)$\\
DFT-MD & 50000 & 108.6(6) & $-53.9581(6)$\\
\hline\\[-6pt]         
$\rho=1.87000$ g$\,$cm$^{-3}$ & & &\\[1pt]
\hline\\[-6pt]                                           
DFT-MD & 2000  & 18.6(1)  & $-54.5887(1)$\\
DFT-MD & 2500  & 20.1(1)  & $-54.5847(1)$\\
DFT-MD & 3500  & 22.3(1)  & $-54.5775(1)$\\
DFT-MD & 5000  & 24.9(1)  & $-54.5659(2)$\\
DFT-MD & 6000  & 26.2(2)  & $-54.5570(3)$\\
DFT-MD & 7500  & 27.5(2)  & $-54.5390(5)$\\
DFT-MD & 10000 & 25.7(3)  & $-54.494(2) $\\
DFT-MD & 15000 & 29.9(2)  & $-54.417(1) $\\
DFT-MD & 20000 & 38.9(2)  & $-54.3538(9)$\\
DFT-MD & 30000 & 63.5(6)  & $-54.2350(8)$\\
DFT-MD & 40000 & 92.3(4)  & $-54.1083(8)$\\
DFT-MD & 50000 & 120(1)   & $-53.9642(6)$\\
\hline\\[-6pt]
$\rho=2.0000$ g$\,$cm$^{-3}$ & & &\\[1pt]
\hline\\[-6pt]                                           
DFT-MD & 2000  & 22.60(8) & $-54.5856(1)$\\
DFT-MD & 2500  & 22.1(1)  & $-54.5802(2)$\\
DFT-MD & 3500  & 25.7(1)  & $-54.5725(1)$\\
DFT-MD & 5000  & 29.2(1)  & $-54.5605(2)$\\
DFT-MD & 6000  & 30.6(2)  & $-54.5512(5)$\\
DFT-MD & 7500  & 29.9(3)  & $-54.529(2) $\\
DFT-MD & 10000 & 29.5(2)  & $-54.4880(8)$\\
DFT-MD & 15000 & 34.1(3)  & $-54.4172(9)$\\
DFT-MD & 20000 & 45.3(4)  & $-54.3537(8)$\\
DFT-MD & 30000 & 72.8(6)  & $-54.2365(7)$\\
DFT-MD & 40000 & 100.7(6) & $-54.1097(8)$\\
DFT-MD & 50000 & 133.2(9) & $-53.9718(7)$\\
\hline\\[-6pt]
$\rho=2.12000$ g$\,$cm$^{-3}$ & & &\\[1pt]
\hline\\[-6pt]
DFT-MD & 2000   & 27.42(6) & $-54.5831(1)$\\
DFT-MD & 2500   & 28.7(1)  & $-54.5790(1)$\\
DFT-MD & 3500   & 31.6(1)  & $-54.5703(2)$\\
DFT-MD & 5000   & 34.2(1)  & $-54.5574(3)$\\
DFT-MD & 6000   & 35.3(2)  & $-54.5475(5)$\\
DFT-MD & 7500   & 35.1(4)  & $-54.526(2) $\\
DFT-MD & 10000  & 33.9(4)  & $-54.482(1) $\\
DFT-MD & 15000  & 38.8(4)  & $-54.4143(8)$\\
DFT-MD & 20000  & 50.9(5)  & $-54.356(1) $\\
DFT-MD & 30000  & 80.0(7)  & $-54.2395(7)$\\
DFT-MD & 40000  & 110.9(6) & $-54.1146(6)$\\
DFT-MD & 50000  & 144(1)   & $-53.9737(5)$\\
\hline\\[-6pt]
$\rho=2.25000$ g$\,$cm$^{-3}$ & & &\\[1pt]
\hline\\[-6pt]
DFT-MD & 2000   & 32.6(1)  & $-54.5798(1)$\\
DFT-MD & 2500   & 34.3(1)  & $-54.5751(1)$\\
DFT-MD & 3500   & 37.4(1)  & $-54.5673(2)$\\
DFT-MD & 5000   & 41.1(1)  & $-54.5540(2)$\\
DFT-MD & 6000   & 40.9(3)  & $-54.5403(6)$\\
DFT-MD & 7500   & 39.3(4)  & $-54.518(1) $\\
DFT-MD & 10000  & 37.1(2)  & $-54.4776(6)$\\
DFT-MD & 15000  & 44.7(5)  & $-54.413(1) $\\
DFT-MD & 20000  & 56.8(4)  & $-54.3568(7)$\\
DFT-MD & 30000  & 89.7(8)  & $-54.2396(9)$\\
DFT-MD & 40000  & 123.2(9) & $-54.1161(8)$\\
DFT-MD & 50000  & 157.9(7) & $-53.9789(4)$\\
\end{longtable}

\begin{longtable}[c]{lrrr}
\hline\\[-6pt]
\multicolumn{4}{c}{Table \ref{long} \emph{Continued}.}\\[1pt]
\hline\\[-6pt]
Method  & T (K) & P (GPa)  &  E (Ha/atom)\\[1pt]
\hline\\[-8pt]

\endfirsthead

\hline\\[-6pt]
\multicolumn{4}{c}{Table \ref{long} \emph{Continued}.}\\[1pt]
\hline\\[-6pt]
Method  & T (K) & P (GPa)  &  E (Ha/atom)\\[1pt]
\hline\\[-8pt]
\endhead

\hline
\endfoot

\hline
\hline
\endlastfoot

$\rho=2.42000$ g$\,$cm$^{-3}$ & & &\\[1pt]
\hline\\[-6pt]
DFT-MD & 2500   & 41.88(8) & $-54.5701(1)$\\
DFT-MD & 3500   & 46.1(1)  & $-54.5618(1)$\\
DFT-MD & 5000   & 49.1(2)  & $-54.5476(3)$\\
DFT-MD & 6000   & 48.2(4)  & $-54.5336(7)$\\
DFT-MD & 7500   & 45.3(5)  & $-54.508(1)$ \\
DFT-MD & 10000  & 43.8(4)  & $-54.4711(7)$\\
DFT-MD & 15000  & 52.9(5)  & $-54.4117(6)$\\
DFT-MD & 20000  & 67.0(5)  & $-54.355(9)$ \\
DFT-MD & 30000  & 102.7(4) & $-54.2410(7)$\\
DFT-MD & 40000  & 139.6(6) & $-54.1185(5)$\\
DFT-MD & 50000  & 179(1)   & $-53.9814(8)$\\

\end{longtable}

\begin{longtable}[c]{lrrrrr}
\hline\\[-6pt]
\multicolumn{6}{c}{Table \ref{long} \emph{Continued}.}\\[1pt]
\hline\\[-6pt]
Method  & T (K) & P (GPa)  &  E (Ha/atom) & P$_{\rm relativistic}$ (GPa) & E$_{\rm relativistic}$ (Ha/atom) \\[1pt]
\hline\\[-8pt]

\endfirsthead

\hline\\[-6pt]
\multicolumn{6}{c}{Table \ref{long} \emph{Continued}.}\\[1pt]
\hline\\[-6pt]
Method  & T (K) & P (GPa)  &  E (Ha/atom) & P$_{\rm relativistic}$ (GPa) & E$_{\rm relativistic}$ (Ha/atom) \\[1pt]
\hline\\[-8pt]
\endhead

\hline
\endfoot

\hline
\hline
\endlastfoot

$\rho=2.52736$ g$\,$cm$^{-3}$ & & &\\[1pt]
\hline\\[-6pt]
DFT-MD  &  2500       &  47.7(1)         & $-54.5671(1)$ & &\\
DFT-MD  &  3500       &  51.0(1)         & $-54.5581(1)$ & &\\
DFT-MD  &  5000       &  54.0(1)         & $-54.5434(2)$ & &\\
DFT-MD  &  6000       &  52.9(5)         & $-54.5280(6)$ & &\\
DFT-MD  &  7500       &  49.8(3)         & $-54.5051(8)$ & &\\
DFT-MD  &  10000      &  49.0(3)         & $-54.4693(5)$ & &\\
DFT-MD  &  15000      &  58.8(5)         & $-54.4095(5)$ & &\\
DFT-MD  &  20000      &  74.1(4)         & $-54.355(7)$  & &\\
DFT-MD  &  30000      &  111.3(5)        & $-54.241(7)$  & &\\
DFT-MD  &  40000      &  147.4(6)        & $-54.1205(3)$ & &\\
DFT-MD  &  50000      &  191.4(9)        & $-53.9856(4)$ & &\\
DFT-MD  &  100000     &  429(1)          & $-53.098(1)$ & &\\
DFT-MD  &  250000     &  1383(3)         & $-49.389(2)$ & &\\
PIMC &  500000     &  3292(12)        & $-41.71(4)$ & &\\
DFT-MD  &  500000     &  3326(5)         & $-42.16(1)$ & &\\
PIMC &  748503     &  5433(11)        & $-34.12(3)$ & &\\
DFT-MD  &  750000     &  5458(7)         & $-34.68(2)$ & &\\
PIMC &  998004     &  7727(9)         & $-24.98(5)$ & &\\
DFT-MD  &  1000000    &  7790(5)         & $-26.71(2)$ & &\\
PIMC &  2020960    &  19466(14)       & 31.69(5) & &\\
PIMC &  4041920    &  45180(12)       & 130.12(4) & &\\
PIMC &  8083850    &  94496(30)       & 290.7(1) & 94496    & 291.2 \\
PIMC &  16167700   &  192028(68)      & 601.0(2) & 192028   & 602.8 \\
PIMC &  49746542   &  595296(310)     & 1879(1)  & 595296   & 1896 \\
PIMC &  99497670   &  1192049(654)    & 3769(2)  & 1192053  & 3837 \\
PIMC &  1034730000 &  12417450(1044)  & 39312(3) & 12417492 & 45648 \\

\end{longtable}

\begin{longtable}[c]{lrrr}
\hline\\[-6pt]
\multicolumn{4}{c}{Table \ref{long} \emph{Continued}.}\\[1pt]
\hline\\[-6pt]
Method  & T (K) & P (GPa)  &  E (Ha/atom)\\[1pt]
\hline\\[-8pt]

\endfirsthead

\hline\\[-6pt]
\multicolumn{4}{c}{Table \ref{long} \emph{Continued}.}\\[1pt]
\hline\\[-6pt]
Method  & T (K) & P (GPa)  &  E (Ha/atom)\\[1pt]
\hline\\[-8pt]
\endhead

\hline
\endfoot

\hline
\hline
\endlastfoot
$\rho=2.75000$ g$\,$cm$^{-3}$ & & &\\[1pt]
\hline\\[-6pt]
DFT-MD & 2500   & 60.9(1)   & $-54.5604(2)$\\
DFT-MD & 3500   & 64.5(1)   & $-54.5497(2)$\\
DFT-MD & 5000   & 64.2(6)   & $-54.5312(8)$\\
DFT-MD & 6000   & 61.4(4)   & $-54.5160(5)$\\
DFT-MD & 7500   & 58.2(2)   & $-54.4941(2)$\\
DFT-MD & 10000  & 59.0(2)   & $-54.4624(5)$\\
DFT-MD & 15000  & 71.8(3)   & $-54.4062(5)$\\
DFT-MD & 20000  & 88.4(4)   & $-54.3519(4)$\\
DFT-MD & 30000  & 130.3(7)  & $-54.2425(4)$\\
DFT-MD & 40000  & 171.7(8)  & $-54.1227(5)$\\
DFT-MD & 50000  & 218.5(5)  & $-53.9871(3)$\\
\hline\\[-6pt]
$\rho=2.90880$ g$\,$cm$^{-3}$ & & &\\[1pt]
\hline\\[-6pt]
DFT-MD & 2500   & 71.1(1)  & $-54.5533(2)$\\
DFT-MD & 3500   & 75.1(2)  & $-54.5434(3)$\\
DFT-MD & 5000   & 70.1(5)  & $-54.5213(6)$\\
DFT-MD & 6000   & 65.6(4)  & $-54.5055(4)$\\
DFT-MD & 7500   & 64.4(3)  & $-54.4863(5)$\\
DFT-MD & 10000  & 68.2(3)  & $-54.4567(4)$\\
DFT-MD & 15000  & 82.5(5)  & $-54.4039(3)$\\
DFT-MD & 20000  & 101.9(4) & $-54.3504(5)$\\
DFT-MD & 30000  & 145.3(5) & $-54.2422(4)$\\
DFT-MD & 40000  & 190.9(7) & $-54.1214(5)$\\
DFT-MD & 50000  & 239.0(8) & $-53.9889(5)$\\
\hline\\[-6pt]
$\rho=3.00000$ g$\,$cm$^{-3}$ & & &\\[1pt]
\hline\\[-6pt]
DFT-MD & 2500   & 77.3(1)  & $-54.5502(1)$\\
DFT-MD & 3500   & 80.3(4)  & $-54.5390(4)$\\
DFT-MD & 5000   & 73.0(5)  & $-54.5157(4)$\\
DFT-MD & 6000   & 70.4(4)  & $-54.5021(3)$\\
DFT-MD & 7500   & 69.0(3)  & $-54.4825(4)$\\
DFT-MD & 10000  & 73.2(2)  & $-54.4544(3)$\\
DFT-MD & 15000  & 89.1(4)  & $-54.4017(5)$\\
DFT-MD & 20000  & 109.2(5) & $-54.3489(3)$\\
DFT-MD & 30000  & 154.6(7) & $-54.2408(4)$\\
DFT-MD & 40000  & 202.3(7) & $-54.1207(3)$\\
DFT-MD & 50000  & 253.0(9) & $-53.9883(5)$\\
\hline\\[-6pt]
$\rho=3.25000$ g$\,$cm$^{-3}$ & & &\\[1pt]
\hline\\[-6pt]
DFT-MD & 2500   & 95.5(1)   & $-54.5396(1)$\\
DFT-MD & 3500   & 89.4(7)   & $-54.5245(2)$\\
DFT-MD & 5000   & 81.2(3)   & $-54.5033(2)$\\
DFT-MD & 6000   & 80.9(3)   & $-54.4912(3)$\\
DFT-MD & 7500   & 83.0(4)   & $-54.4739(3)$\\
DFT-MD & 10000  & 90.4(3)   & $-54.4469(2)$\\
DFT-MD & 15000  & 109.7(3)  & $-54.3968(3)$\\
DFT-MD & 20000  & 131.3(5)  & $-54.3456(4)$\\
DFT-MD & 30000  & 182.2(5)  & $-54.2374(3)$\\
DFT-MD & 40000  & 234.6(9)  & $-54.1195(4)$\\
DFT-MD & 50000  & 288(1)    & $-53.9897(7)$\\
\hline\\[-6pt]
$\rho=3.50000$ g$\,$cm$^{-3}$ & & &\\[1pt]
\hline\\[-6pt]
DFT-MD & 2500   & 97.1(6)   & $-54.5264(2)$\\
DFT-MD & 3500   & 93.0(6)   & $-54.5116(2)$\\
DFT-MD & 5000   & 94.0(3)   & $-54.4931(2)$\\
DFT-MD & 6000   & 96.5(2)   & $-54.4817(1)$\\
DFT-MD & 7500   & 100.5(2)  & $-54.4654(3)$\\
DFT-MD & 10000  & 109.9(3)  & $-54.4403(2)$\\
DFT-MD & 15000  & 133.3(4)  & $-54.3899(5)$\\
DFT-MD & 20000  & 157.8(6)  & $-54.3400(4)$\\
DFT-MD & 30000  & 213.3(6)  & $-54.2332(5)$\\
DFT-MD & 40000  & 269.1(7)  & $-54.1170(6)$\\
DFT-MD & 50000  & 330(1)    & $-53.9865(5)$\\

\end{longtable}

\begin{longtable}[c]{lrrrrr}
\hline\\[-6pt]
\multicolumn{6}{c}{Table \ref{long} \emph{Continued}.}\\[1pt]
\hline\\[-6pt]
Method  & T (K) & P (GPa)  &  E (Ha/atom) & P$_{\rm relativistic}$ (GPa) & E$_{\rm relativistic}$ (Ha/atom) \\[1pt]
\hline\\[-8pt]

\endfirsthead

\hline\\[-6pt]
\multicolumn{6}{c}{Table \ref{long} \emph{Continued}.}\\[1pt]
\hline\\[-6pt]
Method  & T (K) & P (GPa)  &  E (Ha/atom) & P$_{\rm relativistic}$ (GPa) & E$_{\rm relativistic}$ (Ha/atom) \\[1pt]
\hline\\[-8pt]
\endhead

\hline
\endfoot

\hline
\hline
\endlastfoot

$\rho=3.70586$ g$\,$cm$^{-3}$ & & &\\[1pt]
\hline\\[-6pt]				
DFT-MD  &  2500       & 92.3(6)         & $-54.5207(3)$ & & \\
DFT-MD  &  3500       & 100.1(7)        & $-54.5038(2)$ & & \\
DFT-MD  &  5000       & 107.1(3)        & $-54.4859(2)$ & & \\
DFT-MD  &  6000       & 111.2(2)        & $-54.4746(1)$ & & \\
DFT-MD  &  7500       & 117.5(3)        & $-54.4588(2)$ & & \\
DFT-MD  &  10000      & 129.0(3)        & $-54.4339(3)$ & & \\
DFT-MD  &  15000      & 155.2(4)        & $-54.3847(3)$ & & \\
DFT-MD  &  20000      & 182.0(5)        & $-54.3345(3)$ & & \\
DFT-MD  &  30000      & 240.1(6)        & $-54.2293(4)$ & & \\
DFT-MD  &  40000      & 301.8(9)        & $-54.1128(5)$ & & \\
DFT-MD  &  50000      & 364.1(8)        & $-53.9859(4)$ & & \\
DFT-MD  &  100000     & 736(4)          & $-53.156(2)$ & & \\
PIMC &  250000     & 2041(25)        & $-49.03(5)$ & & \\
DFT-MD  &  250000     & 2128(5)         & $-49.674(3)$ & & \\
PIMC &  500000     & 4906(17)        & $-42.23(3)$ & & \\
DFT-MD  &  500000     & 4919(9)         & $-42.82(1)$ & & \\
PIMC &  748503     & 7900(19)        & $-35.09(4)$ & & \\
DFT-MD  &  750000     & 8002(14)        & $-35.50(3)$ & & \\
PIMC &  998004     & 11155(20)       & $-26.47(5)$ & & \\
DFT-MD  &  1000000    & 11388(25)       & $-27.67(7)$ & & \\
PIMC &  2020960    & 27900(20)       & 26.71(5) & & \\
PIMC &  4041920    & 65381(16)       & 125.62(4) & & \\
PIMC &  8083850    & 138020(36)      & 288.00(9) & 138020     & 288.4\\
PIMC &  16167700   & 280934(109)     & 598.5(2)  & 280934     & 600.3\\
PIMC &  49746542   & 872493(429)     & 1877.3(9) & 872492     & 1894.5\\
PIMC &  99497670   & 1748778(888)    & 3770(2)   & 1748783    & 3838\\
PIMC &  1034730000 & 18205150(1897)  & 39306(4)  & 18205208   & 45642\\
\hline\\[-6pt]
$\rho=7.82952$ g$\,$cm$^{-3}$ & & &\\[1pt]
\hline\\[-6pt]				
DFT-MD  &  10000      &  1116(2)         & $-54.183(1)$ & &\\
DFT-MD  &  50000      &  1661(4)         & $-53.757(2)$ & &\\
DFT-MD  &  100000     &  2413(4)         & $-53.050(1)$ & &\\
PIMC &  250000     &  5385(54)        & $-48.80(5)$ & &\\
DFT-MD  &  250000     &  5342(9)         & $-49.964(4)$ & &\\
PIMC &  500000     &  10758(41)       & $-43.28(4)$ & &\\
DFT-MD  &  500000     &  11150(15)       & $-43.61(1)$ & &\\
PIMC &  748503     &  17030(42)       & $-36.49(4)$ & &\\
DFT-MD  &  750000     &  17366(27)       & $-36.81(2)$ & &\\
PIMC &  998004     &  23651(30)       & $-28.73(3)$ & &\\
DFT-MD  &  1000000    &  24111(42)       & $-29.55(5)$ & &\\
PIMC &  2020960    &  56615(47)       & 17.53(5) & &\\
PIMC &  4041920    &  134303(34)      & 115.31(4) & &\\
PIMC &  8083850    &  288195(112)     & 280.5(1) & 288194    & 280.9\\
PIMC &  16167700   &  587988(204)     & 589.4(2) & 587987    & 591.2\\
PIMC &  49746542   &  1841498(1014)   & 1873(1)  & 1841498  & 1890\\
PIMC &  99497670   &  3694944(1982)   & 3786(2)  & 3694944  & 3837\\
PIMC &  1034730000 &  38463564(3144)  & 39305(3) & 38463672 & 45641\\
\hline\\[-6pt]
$\rho=13.9462$ g$\,$cm$^{-3}$ & & &\\[1pt]
\hline\\[-6pt]				
DFT-MD  &  10000      &  4526(12)       & $-53.608(2)$ & &\\
DFT-MD  &  50000      &  5734(5)        & $-53.108(1)$ & &\\
DFT-MD  &  100000     &  7032(8)        & $-52.473(2)$ & &\\
PIMC &  250000     &  13170(99)      & $-47.68(5)$ & &\\
DFT-MD  &  250000     &  12028(9)       & $-49.696(3)$ & &\\
PIMC &  500000     &  20971(69)      & $-43.59(4)$ & &\\
DFT-MD  &  500000     &  22039(20)      & $-43.778(8)$ & &\\
PIMC &  748503     &  32093(77)      & $-36.97(4)$ & &\\
DFT-MD  &  750000     &  32787(23)      & $-37.34(1)$ & &\\
PIMC &  998004     &  43128(68)      & $-29.99(4)$ & &\\
DFT-MD  &  1000000    &  44391(82)      & $-30.41(5)$ & &\\
PIMC &  2020960    &  99060(65)      & 11.48(4)  & &\\
PIMC &  4041920    &  233718(56)     & 106.13(4) & &\\
PIMC &  8083850    &  508193(153)    & 273.4(1)  & 508193  & 273.9\\
PIMC &  16167700   &  1049077(357)   & 587.9(2)  & 1049077 & 589.7\\
PIMC &  49746542   &  3277237(1514)  & 1869.6(9) & 3277237 & 1886.8\\
PIMC &  99497670   &  6577836(2765)  & 3764(2)   & 6577836  & 3832\\
PIMC &  1034730000 &  68515111(6897) & 39305(4)  & 68515322 & 45641\\
\end{longtable}